\documentclass{IEEEtran}
\usepackage{mathrsfs}
\usepackage{algorithm}
\usepackage{algorithmic}
\usepackage{cite}
\usepackage{color}
\usepackage{psfrag}
\usepackage{subcaption}
\usepackage{amssymb}
\usepackage{epsfig}
\usepackage{pifont}
\usepackage{amsmath}
\usepackage{array}
\usepackage{pifont}
\usepackage{amsfonts}
\usepackage{fancyhdr}
\usepackage{graphics,eepic,epic}
\usepackage{latexsym}
\usepackage{euscript}
\usepackage{graphics,eepic,epic,psfrag}
\usepackage{graphicx,bm,dsfont}
\usepackage{xcolor}
\usepackage{float}
\usepackage{flafter}

\usepackage{enumerate}

\usepackage{hyperref}
\hypersetup{
	colorlinks=true,
	linkcolor=blue,
	citecolor=blue
}

\newtheorem{theorem}{Theorem}
\newtheorem{proposition}{Proposition}

\newtheorem{assumption}{Assumption}

\newcommand{\op}[1]{\mathrm{#1}}

\newcommand{\V}{\mathbf{v}}
\newcommand{\lip}{\textit{\L}}
\newcommand{\qam}{\left\{ \pm \frac{\sqrt{2}}{2} \pm j \frac{\sqrt{2}}{2} \right\}}

\usepackage{tikz}
\usetikzlibrary{shapes,snakes,plotmarks}
\usepackage{pgfplots}
\usetikzlibrary{arrows.meta}

\usepackage{makecell}
\usepackage{adjustbox}

\begin{document}
\title{Covariance-Based Device Activity Detection with Massive MIMO for Near-Field Correlated Channels}
\author{Ziyue Wang, 
		Yang Li,~\IEEEmembership{Member,~IEEE,}
        Ya-Feng Liu,~\IEEEmembership{Senior~Member,~IEEE,}
        and Junjie Ma 
\thanks{
Received 11 November 2024; revised 24 March 2025 and 9 July 2025; accepted 8 September 2025.
The work of Ziyue Wang and Ya-Feng Liu was supported in part by the National Natural Science Foundation of China (NSFC) under Grant 12371314 and Grant 12021001.
The work of Yang Li was supported in part by Guangdong Basic and Applied Basic Research Foundation under Grant 2025A1515011658, and in part by the NSFC under Grant 62101349.
The work of Junjie Ma was supported in part by the NSFC under Grant 62571526.
An earlier version of this paper was presented in part at the 2025 IEEE/CIC International Conference on Communications in China~\cite{wang2025covariance}.
\textit{(Corresponding author: Ya-Feng Liu.)}

Ziyue Wang and Junjie Ma are with the State Key Laboratory of Mathematical Sciences, Academy of Mathematics and Systems Science, Chinese Academy of Sciences, Beijing 100190, China (e-mails: ziyuewang@lsec.cc.ac.cn, majunjie@lsec.cc.ac.cn).

Yang Li is with the School of Computing and Information Technology, Great Bay University, Dongguan 523000, China, and also with Dongguan Key Laboratory for Intelligence and Information Technology, Dongguan 523000, China (e-mail: liyang@gbu.edu.cn).

Ya-Feng Liu is with the Ministry of Education Key Laboratory of Mathematics and Information Networks, School of Mathematical Sciences, Beijing University of Posts and Telecommunications, Beijing 102206, China (e-mail: yafengliu@bupt.edu.cn).
}
}

\maketitle

\begin{abstract}
This paper studies the device activity detection problem in a massive multiple-input multiple-output (MIMO) system for near-field communications (NFC).
In this system, active devices transmit their signature sequences to the base station (BS), which detects the active devices based on the received signal.
In this paper, we model the near-field channels as correlated Rician fading channels and formulate the device activity detection problem as a maximum likelihood estimation (MLE) problem. Compared to the traditional uncorrelated channel model, the correlation of channels complicates both algorithm design and theoretical analysis of the MLE problem.
On the algorithmic side, we present the classical exact coordinate descent (CD) algorithm for solving the MLE problem, which suffers from numerical instability when applied to correlated channels.
We propose a computationally efficient inexact CD algorithm by approximating the objective function, which approximately solves the one-dimensional subproblem and improves both computational efficiency and numerical stability.
Additionally, we analyze the detection performance of the MLE problem under correlated channels by comparing it with the case of uncorrelated channels.
The analysis shows that when the overall number of devices $\bm{N}$ is large or the signature sequence length $\bm{L}$ is small, the detection performance of MLE under correlated channels tends to be better than that under uncorrelated channels.
Conversely, when $\bm{N}$ is small or $\bm{L}$ is large, MLE performs better under uncorrelated channels than under correlated ones.
Finally, we study the MLE model in the joint device activity and data detection context.
Simulation results demonstrate the computational performance of the presented algorithms and verify the correctness of the analysis.
\end{abstract}
\begin{IEEEkeywords}
Coordinate descent (CD) algorithm, device activity detection, massive random access, massive multiple-input multiple-output (MIMO), near-field communications (NFC).
\end{IEEEkeywords}
\IEEEpeerreviewmaketitle

\section{Introduction}

Massive machine-type communication (mMTC) plays a crucial role in fifth-generation (5G) and beyond cellular networks~\cite{bockelmann2016massive}.
In mMTC, a key challenge is to handle massive random access, in which a large number of sporadically active devices are connected to the network~\cite{chen2021massive}.
This challenge can be addressed during the \textit{activity detection} phase, in which each device is preassigned a unique nonorthogonal signature sequence, and the active devices transmit their signature sequences to the base station~(BS).
The BS then detects the active devices by identifying which sequences are transmitted based on the received signal~\cite{liu2018sparse}.

There are generally two mathematical approaches for device activity detection.
The first one is the compressed sensing (CS)-based approach, which jointly estimates the device activities and instantaneous channel state information (CSI) by exploiting the sporadic nature of device activities~\cite{gao2024compressive}.
The second one is the covariance-based approach, which only estimates the device activities by taking advantage of the statistical CSI~\cite{haghighatshoar2018improved}.
It formulates the activity detection problem as a maximum likelihood estimation (MLE) problem, which is referred to as the covariance-based approach because the MLE formulation only depends on the sample covariance matrix of the received signal.
Studies have shown that the covariance-based approach not only outperforms the CS-based approach but also correctly detects a much larger number of active devices in massive multiple-input multiple-output (MIMO) systems~\cite{fengler2021non,chen2022phase,wang2023covariance}.

The device activity detection problem has been extensively studied with respect to theoretical analysis, algorithm design, and practical issues, with most existing works focusing on classic far-field uncorrelated Rayleigh fading channels~\cite{fengler2021non,chen2022phase,wang2023covariance,chen2019covariance,wang2022covariance,li2023asynchronous,liu2024grant,liu2024survey
}.
However, with emerging trends such as extremely large-scale MIMO (XL-MIMO) and tremendously high frequencies in sixth-generation (6G) networks, the traditional far-field assumption is no longer accurate~\cite{liu2023near,lu2024tutorial}.
This is because the Rayleigh distance, which separates the near-field and far-field regions, is proportional to the square of the antenna aperture and signal frequency.
Consequently, many devices might be within the near-field region.
Therefore, it is necessary to adopt the more practical near-field channel model for device activity detection.

In this paper, we study the covariance-based activity detection problem under the spatially correlated Rician fading channel, where the deterministic line-of-sight (LoS) components and correlation matrices for the near-field devices are derived from near-field spherical wave propagation~\cite{liu2023near,lu2024tutorial,dong2022near,dong2024characterizing}.
This setup makes both algorithm design and detection performance analysis more challenging than those under traditional uncorrelated Rayleigh fading channels, primarily due to the channel correlation.
For spatially uncorrelated channels, the MLE formulation relies on the marginal distributions of each column of the received signal, as these columns are independently distributed~\cite{fengler2021non}.
However, in the case of spatially correlated channels, different columns of the received signal are dependent with each other and it is necessary to consider their joint distribution.
The joint distribution complicates the MLE problem because its dimension is $LM,$ which is much larger than that of the marginal distribution (i.e., $L$), where $L$ is the signature sequence length and $M$ is the number of antennas. This is particularly relevant in the context of massive MIMO.
Moreover, since the existing detection performance analyses~\cite{fengler2021non,chen2022phase,wang2023covariance} assume that different columns of the received signal are independent and identically distributed~(i.i.d.), they cannot be directly applied to the case of spatially correlated channels.

\subsection{Related Works}

The CS-based approach is an important line of research in device activity detection under uncorrelated Rayleigh fading channels, which takes advantage of the sporadic nature of the device activity pattern.
Existing sparse recovery methods for solving the CS-based activity detection problem include approximate message passing (AMP)~\cite{chen2018sparse,liu2018massive,senel2018grant,bian2023joint,zhu2024cooperative} and group least absolute shrinkage and selection operator (LASSO)~\cite{li2022dynamic,sun2023joint,marata2023joint}.

Another important line of research in device activity detection is the covariance-based approach, which exploits the statistical information of the channel and formulates the activity detection problem as an MLE problem.
From an algorithmic perspective, the coordinate descent (CD) algorithm~\cite{haghighatshoar2018improved} and its accelerated variants~\cite{dong2022faster,wang2023covariance,zhu2024rethinking} are the most popular and efficient algorithms for solving the MLE problem.
Other algorithms include the expectation maximization/minimization (EM), also known as sparse Bayesian learning~\cite{wipf2007empirical}, the sparse iterative covariance-based estimation (SPICE)~\cite{stoica2011spice}, and gradient descent~\cite{wang2021efficient}. From a theoretical perspective, the covariance-based approach has been demonstrated to have a quadratic scaling law, which enables it to detect many more active devices than the CS-based approach~\cite{fengler2021non,chen2022phase,wang2023covariance}.
However, the above algorithms and theories all rely on the assumption of uncorrelated Rayleigh fading channels and are not directly applicable to correlated Rayleigh/Rician fading channels.

The CS-based approach under correlated Rayleigh fading channels has been studied in~\cite{djelouat2021user,bai2022activity,rajoriya2023joint,cheng2021orthogonal,djelouat2021joint,djelouat2022spatial,djelouat2024hierarchical,xie2024massive}.
Specifically, by utilizing the channel distribution information, the AMP algorithm~\cite{djelouat2021user,bai2022activity,rajoriya2023joint} and orthogonal AMP (OAMP) algorithm~\cite{ma2017orthogonal,cheng2021orthogonal} are proposed for solving the device activity detection problem.
The works \cite{djelouat2021joint,djelouat2022spatial} study the problem formulation of the CS-based activity detection and utilize the alternating direction method of multipliers (ADMM) algorithm for solving it.
Additionally, \cite{djelouat2024hierarchical} proposes a Bayesian framework for CS-based activity detection with unknown correlation matrices, and \cite{xie2024massive} proposes a turbo-type compressive sampling matching pursuit (Turbo-CoSaMP) algorithm for solving the device activity problem by exploiting the structured sparsity of the near-field channel.

However, there are few studies on the covariance-based approach under correlated Rayleigh or uncorrelated Rician fading channels.
Three works~\cite{xie2020massive,lin2024intelligent,bai2024unified} investigate the covariance-based approach under correlated Rayleigh fading channels.
In \cite{xie2020massive}, the MLE formulation and algorithm are studied under the assumption that the eigenvectors for the correlation matrices of all devices are identical.
In \cite{lin2024intelligent}, the covariance-based approach is applied to intelligent reflecting surface (IRS)-aided activity detection, and a deep unfolding approach is proposed for solving the MLE problem.
The work~\cite{bai2024unified} studies the covariance-based approach when the channels are correlated within time-frequency blocks.
Moreover, under uncorrelated Rician fading channels, the problem formulation and algorithms of the covariance-based approach are studied in~\cite{liu2023mle} and \cite{tian2022massive}.
In this paper, we study the algorithm design and theoretical analysis of the covariance-based approach considering correlated Rician fading channels.

The covariance-based approach for device activity detection has been applied in various practical scenarios, including the joint activity and data detection problem~\cite{chen2019covariance,gao2023energy,lin2022sparsity}, the asynchronous system with joint activity and delay detection~\cite{wang2022covariance,li2023asynchronous}, the multi-cell system~\cite{wang2023covariance,chen2021sparse}, the cell-free system~\cite{zhang2024activity,lin2024communication}, the unsourced random access scenario~\cite{fengler2021non}, the system with frequency offsets~\cite{li2019activity,liu2023mle}, the system with low-resolution analog-to-digital converters~\cite{wang2023device}, and the orthogonal frequency division multiplexing (OFDM)-based massive grant-free access scheme~\cite{jia2023statistical}.
Moreover, the deep learning approach is applied to device activity detection~\cite{li2023heterogeneous}.
Finally, the covariance-based approach can be used in a three-phase protocol~\cite{kang2021minimum,kang2022scheduling} to first identify active devices and then schedule them in orthogonal transmission slots for channel estimation.

\subsection{Main Contributions}

This paper investigates the covariance-based activity detection problem in massive MIMO systems under spatially correlated Rician fading channels.
The device activity detection problem is formulated as an MLE problem.
The correlation of channels complicates both the algorithm design and theoretical analysis of the MLE problem.
First, due to the channel correlation, when applying the CD algorithm, there is no closed-form solution to the one-dimensional optimization subproblem, and the dimension of the covariance matrix in the MLE formulation is proportional to the number of antennas, which is significantly different from the case where the channels are uncorrelated.
Second, compared with the uncorrelated channel case, the channel correlation makes the columns of the received signal no longer i.i.d., which challenges the theoretical analysis of MLE's detection performance.
The main contributions of the paper are twofold.

On the algorithmic side, we introduce an exact CD algorithm for solving the MLE problem, which is a direct extension of the CD algorithm designed for the uncorrelated channel case~\cite{haghighatshoar2018improved}. It exactly solves the one-dimensional optimization subproblem by using matrix eigenvalue decomposition and polynomial root-finding techniques.
However, this algorithm suffers from numerical instability and can fail to converge when the channel correlation matrix has a high rank. To address this issue, we propose an inexact CD algorithm, which primarily relies on Taylor expansion to approximate selected terms in the objective function. This approach provides improved numerical stability and reduced per-coordinate update complexity.
Additionally, we study a special case where only a few devices’ channels are correlated with a low-rank correlation matrix, while the channels of many other devices are uncorrelated. We demonstrate that exploiting the low-rank properties of the correlation matrices significantly reduces the complexity of both the exact CD and inexact CD algorithms.

On the analytical side, we analyze the detection performance of the MLE problem under correlated channels by comparing it with the case of uncorrelated channels.
This comparison is examined from two perspectives: signal statistical dimension and identifiability of the true activity.
From the signal statistical dimension perspective, due to the low-rank property of the near-field spatial correlation matrix, the detection performance of MLE under correlated channels tends to be worse than that under uncorrelated channels when the overall number of devices $N$ is small or the signature sequence length $L$ is large.
Conversely, from the perspective of identifiability, correlated channels make the identifiability of the true activity more likely to hold than uncorrelated channels, suggesting that MLE performs better in correlated channels than in uncorrelated ones when $N$ is large or $L$ is small.

In addition, we study the joint device activity and data detection problem, where each active device wishes to send a few bits of information to the BS. In this scenario, each device is associated with multiple signature sequences and can embed a small number of information bits. We propose the MLE model for this scenario. Simulation results show that transmitting several bits of information does not significantly deteriorate the detection performance of the MLE model, especially when the length of the signature sequence is large.

\subsection{Paper Organization and Notations}

The rest of the paper is organized as follows.
Section~\ref{sec:system} introduces the system model and the MLE formulation.
Section~\ref{sec:algorithm} presents the efficient CD algorithms for solving the MLE problem.
In Section~\ref{sec:analysis}, we analyze the difference in detection performance between correlated and uncorrelated channels.
In Section~\ref{sec:data-detection}, we study the MLE model in the joint device activity and data detection context.
Simulation results are provided in Section~\ref{sec:simulation}.
Finally, we conclude the paper in Section~\ref{sec:conclusion}.

In this paper, lower-case, boldface lower-case, and boldface upper-case letters denote scalars, vectors, and matrices, respectively.
Calligraphic letters denote sets.
Superscripts $(\cdot)^{H},$ $(\cdot)^{T},$ $(\cdot)^{*},$ and $(\cdot)^{-1},$ denote conjugate transpose, transpose, conjugate, and inverse, respectively.
Additionally, $j$ denotes the imaginary unit, $\bm{I}_M$ denotes the identity matrix of dimension $M\times  M,$ $\mathbb{E}[\cdot]$ denotes expectation, $\op{tr}(\bm{X})$ denotes the trace of $\bm{X},$ $\op{diag}(x_1,x_2,\ldots,x_n)$ (or $\op{diag}(\bm{X}_1,\bm{X}_2,\ldots,\bm{X}_n)$) denotes a (block) diagonal matrix formed by $x_1,x_2,\ldots,x_n$ (or $\bm{X}_1,\bm{X}_2,\ldots,\bm{X}_n$), $\triangleq$ denotes definition, $|\cdot|$ denotes the determinant of a matrix, or the (coordinate-wise) absolute operator, $\|\bm{x}\|_p$ denotes the $\ell_p$-norm of $\bm{x},$ and $\otimes$ denotes the Kronecker product. Finally, $\mathcal{CN}(\boldsymbol\mu, \bm{\Sigma})$ denotes a complex Gaussian distribution with mean $\boldsymbol\mu$ and covariance $\bm{\Sigma}.$

\section{System Model and Problem Formulation}
\label{sec:system}

\subsection{System Model}
\label{subsec:system}

\begin{figure}[t]
	\centering
	\begin{adjustbox}{width=0.97\columnwidth,center}
		\begin{tikzpicture}
			\draw[fill=gray!10,draw=black,thick] (0,0) circle [x radius=100pt,y radius=50pt];
			\draw[fill=yellow!10,draw=black,thick] (0,0) circle [x radius=30pt,y radius=25pt];
			\node[fill=orange, regular polygon, regular polygon sides=3,inner sep=2pt] at (0,0) {};
			\draw[fill=blue] (-0.8,1.2) circle [radius=3pt];
			\draw[fill=blue] (-0.6,-1.2) circle [radius=3pt];
			\draw[fill=blue] (1.5,0.1) circle [radius=3pt];
			\draw[fill=blue] (0.15,0.5) circle [radius=3pt];
			\draw[fill=red] (-0.5,-0.25) circle [radius=3pt];
			\draw[fill=blue] (-1.2,0.4) circle [radius=3pt];
			\draw[fill=blue] (0.5,-0.1) circle [radius=3pt];
			\draw[fill=red] (1,1.5) circle [radius=3pt];
			\draw[fill=red] (-1,-0.8) circle [radius=3pt];
			\draw[fill=blue] (-3,-0.5) circle [radius=3pt];
			\draw[fill=blue] (-2.5,1) circle [radius=3pt];
			\draw[fill=blue] (2.7,0.7) circle [radius=3pt];
			\draw[fill=red] (2.4,-0.2) circle [radius=3pt];
			\draw[fill=red] (0.7,-1.5) circle [radius=3pt];
			\node[fill=orange,regular polygon, regular polygon sides=3,inner sep=2pt] at (-3.8,-2.5) {};
			\node at (-2.6,-2.5) [text opacity=1] {Base station};
			\draw[fill=red] (-1.2,-2.5) circle [radius=3pt];
			\node at (0.0,-2.5) [text opacity=1] {Active device};
			\draw[fill=blue] (1.4,-2.5) circle [radius=3pt];
			\node at (2.8,-2.5) {Inactive device};
			\node at (-2,-2.1) {Far-field region};
			\node at (2,-2.1) {Near-field region};
			\draw[->, thick] (0.3,-0.5) to[out=0,in=90] (2,-1.85);
			\draw [->, thick] (-1.5,-1) to[out=180,in=90] (-2,-1.85);
		\end{tikzpicture}
	\end{adjustbox}
	\caption{System model for massive random access with near-field and far-field devices.}
	\label{fig:system-model}
\end{figure}

We consider an uplink single-cell massive MIMO system consisting of a BS equipped with an $M$-antenna uniform linear array with half-wavelength antenna spacing.
In the cell, there are $N$ single-antenna devices, and only $K\ll N$ devices are active during any coherence interval due to the sporadic traffic.
We use $a_{n}$ to denote the binary variable with $a_{n}=1$ for active and $a_{n}=0$ for inactive devices.
We consider the case where some devices are located in the near-field region of the BS, and others in the far-field region, as shown in Fig.~\ref{fig:system-model}.
The radius of the near-field region is indicated by the Rayleigh distance, which is given by $2D^2/\lambda,$ where $D$ and $\lambda$ denote the antenna aperture and the wavelength, respectively~\cite{liu2023near}.
Notice when the carrier frequency is set as $3$\,GHz, and the number of antennas is set as $32,$ $64,$ and $128,$ the corresponding Rayleigh distances are approximately $50$\,m, $200$\,m, and $800$\,m, respectively.
Consequently, the number of near-field devices can be large due to the substantial Rayleigh distance.

We assume that both the deterministic LoS channel and statistical multipath channels exist, and the channel between device $n$ and the BS can be modeled as the following Rician fading model:
\begin{equation}\label{eq:channel-distribution}
	\bm{h}_{n} \sim \mathcal{CN}\big(\overline{\bm{h}}_n,\bm{R}_n\big),
\end{equation}
where $\overline{\bm{h}}_n \in \mathbb{C}^M$ is the LoS channel between device $n$ and the BS. 
For a near-field device $n,$ the mean vector $\overline{\bm{h}}_n$ can be expressed as \cite[Eq.~(133)]{liu2023near}:
\begin{equation}\label{eq:los-channel}
	\big(\overline{\bm{h}}_n\big)_m = \beta_n \exp \left( -j \frac{2\pi}{\lambda} (d_{n,m} - d_{n,0}) \right), \, m = 1,2,\ldots, M,
\end{equation}
where $\beta_n$ denotes the channel gain between device $n$ and the BS, $d_{n,m}$ is the distance between the device and the $m$-th antenna, and $d_{n,0}$ is the distance between the device and the midpoint of the antenna array.
Furthermore, the correlation matrix $\bm{R}_n \in \mathbb{C}^{M \times M}$ is given by \cite[Eq.~(134)]{liu2023near}:
\begin{equation}\label{eq:R-structure}
	\bm{R}_n = \sum_{\ell=1}^{\bar{\ell}_n} \sigma_{\ell}^2 \, |\beta_{n,\ell}|^2 \, \overline{\bm{h}}_{\ell} \big(\overline{\bm{h}}_{\ell}\big)^H,
\end{equation}
where $\bar{\ell}_n$ is the number of scatterers between device $n$ and the BS, $\sigma_{\ell}^2$ is the intensity attenuation caused by the $\ell$-th scatterer, $\beta_{n,\ell}$ denotes the channel gain between the $\ell$-th scatterer and the device $n,$ and $\overline{\bm{h}}_{\ell}$ is the channel vector between the $\ell$-th scatterer and the BS with a similar structure to~\eqref{eq:los-channel}, i.e., $(\overline{\bm{h}}_{\ell})_m = \beta_{\ell} \exp \left( -j \frac{2\pi}{\lambda} (d_{\ell,m} - d_{\ell,0}) \right),$ where $\beta_{\ell},$ $d_{\ell,m},$ and $d_{\ell,0}$ denote the corresponding channel gain and distances.
Notice that $\mathrm{rank}(\bm{R}_n) \le \bar{\ell}_n,$ and thus $\bm{R}_n$ is a low-rank matrix when the number of scatterers $\bar{\ell}_n \ll M.$
For a far-field device $n,$ the vector $\overline{\bm{h}}_n$ can be represented by the conventional plane wave model, and the correlation matrix can be modeled as $\bm{R}_n = g_n \bm{I}_M,$ where $g_n>0$ is the large-scale fading coefficient~\cite{liu2023mle}.

Each device $n$ is preassigned a unique signature sequence $\bm{s}_{n}\in\mathbb{C}^{L}$ with length $L$ for activity detection.
During the uplink pilot stage, all active devices synchronously transmit their signature sequences to the BS as random access requests. Following this, the received signal at the BS can be expressed as~\cite{haghighatshoar2018improved,fengler2021non,chen2022phase}
\begin{equation}\label{eq:sys}
	\bm{Y} = \sum_{n=1}^{N}a_{n}\bm{s}_{n}\bm{h}_{n}^T  + \bm{W},
\end{equation}
where $\bm{W} \in \mathbb{C}^{L \times M}$ is the additive i.i.d. white Gaussian noise with variance $\sigma_w^2.$ 
Denote $\bm{a} = [a_1, a_2, \ldots, a_N]^T \in \mathbb{R}^{N}$ as the activity indicator vector.

\subsection{Problem Formulation}
\label{subsec:prob-form}

The task of device activity detection is to estimate $\bm{a}$ from the received signal $\bm{Y}.$
Notice that the Rician fading components and noises are both Gaussian.
Therefore, the received signal $\bm{Y}$ in \eqref{eq:sys} follows a Gaussian distribution.
Since the columns of $\bm{Y}$ are generally not i.i.d., we need to consider the joint distribution of all its columns, i.e., the distribution of its vectorization:
\begin{equation}
	\bm{y} \triangleq \operatorname{vec}(\bm{Y}) = \sum_{n=1}^{N}a_{n} \bm{h}_{n} \otimes \bm{s}_{n} + \bm{w},
\end{equation}
where $\bm{w} = \operatorname{vec}(\bm{W}).$
For a given $\bm{a},$ $\bm{y}$ is a Gaussian vector, i.e., $\bm{y}\sim\mathcal{CN}(\overline{\bm{y}}_{\bm{a}},\bm{\Sigma}_{\bm{a}}),$ where the mean is
\begin{equation}\label{eq:def-y-bar}
	\overline{\bm{y}}_{\bm{a}} = \mathbb{E} [ \bm{y} ]  = \sum_{n=1}^{N}a_{n} \overline{\bm{h}}_{n} \otimes \bm{s}_{n}, 
\end{equation}
and the covariance matrix is given by
\begin{align}\label{eq:def-cov}
	\bm{\Sigma}_{\bm{a}} & = \mathbb{E}[ (\bm{y}-\overline{\bm{y}}_{\bm{a}}) (\bm{y}-\overline{\bm{y}}_{\bm{a}})^H ] \nonumber \\
	& = \sum_{n=1}^{N} a_{n} \mathbb{E}\left[ \left(\left(\bm{h}_{n}-\overline{\bm{h}}_{n}\right) \otimes \bm{s}_{n}\right)  \left(\left(\bm{h}_{n}-\overline{\bm{h}}_{n}\right) \otimes \bm{s}_{n}\right)^H \right] \nonumber \\
	& \qquad \qquad \qquad \qquad \qquad \qquad \qquad \qquad \qquad + \sigma_w^2 \bm{I}_{LM}  \nonumber \\
	& = \sum_{n=1}^{N} a_{n} \mathbb{E}\left[ \left(\left(\bm{h}_{n}-\overline{\bm{h}}_{n}\right) \left(\bm{h}_{n}-\overline{\bm{h}}_{n}\right)^H \right) \otimes \left(\bm{s}_{n} \bm{s}_{n}^H\right) \right] \nonumber \\
	& \qquad \qquad \qquad \qquad \qquad \qquad \qquad \qquad \qquad + \sigma_w^2 \bm{I}_{LM}  \nonumber \\
	& = \sum_{n=1}^{N} a_{n} \bm{R}_n \otimes \left(\bm{s}_{n} \bm{s}_{n}^H\right) + \sigma_w^2 \bm{I}_{LM}.
\end{align}
Therefore, the likelihood function of $\bm{y}$ can be expressed as
\begin{equation}
	p( \bm{y} \,|\, \bm{a} ) = \frac{1}{| \pi \bm{\Sigma}_{\bm{a}} |} \exp \left( -  \left(\bm{y}-\overline{\bm{y}}_{\bm{a}}\right)^H \bm{\Sigma}_{\bm{a}}^{-1}  \left(\bm{y}-\overline{\bm{y}}_{\bm{a}}\right) \right).
\end{equation}
The MLE problem can be formulated as the minimization of the negative log-likelihood function $- \log p(\bm{y}\,|\,\bm{a}),$ which can be expressed as
\begin{subequations}\label{eq:mle-binary}
	\begin{alignat}{2}
		&\underset{\bm{a}}{\operatorname{minimize}}    &\quad& \log|\bm{\Sigma}_{\bm{a}}| +   \left(\bm{y}-\overline{\bm{y}}_{\bm{a}}\right)^H \bm{\Sigma}_{\bm{a}}^{-1} \left(\bm{y}-\overline{\bm{y}}_{\bm{a}}\right)\\
		&\operatorname{subject\,to} &      &a_{n} \in \{0,1\}, \,\forall \, n=1,2,\ldots,N. \label{eq:mle-binary-b}
	\end{alignat}
\end{subequations}
Problem~\eqref{eq:mle-binary} is computationally intractable due to its binary constraints~\eqref{eq:mle-binary-b}.
To make the problem more computationally tractable, we relax the constraint $a_{n} \in \{0,1\}$ to $a_{n} \in [0,1]$ as in \cite{fengler2021non,chen2022phase,wang2023covariance}, and obtain the following relaxed problem:
\begin{subequations}\label{eq:mle}
	\begin{alignat}{2}
		&\underset{\bm{a}}{\operatorname{minimize}}    &\quad& \log|\bm{\Sigma}_{\bm{a}}| +   \left(\bm{y}-\overline{\bm{y}}_{\bm{a}}\right)^H \bm{\Sigma}_{\bm{a}}^{-1} \left(\bm{y}-\overline{\bm{y}}_{\bm{a}}\right)\\
		&\operatorname{subject\,to} &      &a_{n} \in [0,1], \,\forall \, n=1,2,\ldots,N.
	\end{alignat}
\end{subequations}
When solving the MLE problem~\eqref{eq:mle} for device activity detection, we only need to know the mean $\overline{\bm{h}}_n$ and the correlation matrix $\bm{R}_n$ of each device's channel, without requiring the knowledge of the channel vector $\bm{h}_n,$ i.e., the channel state.
Notice that $(\cdot)^T$ appears in \eqref{eq:sys} because of the system model~\cite{haghighatshoar2018improved,fengler2021non,chen2022phase}, and $(\cdot)^H$ appears in \eqref{eq:mle} due to the probability distribution function of the complex Gaussian distribution~\cite{andersen1995multivariate}.

In this paper, we focus on the algorithm design and performance analysis for the MLE problem~\eqref{eq:mle}.
Notice that the dimension of the covariance matrix $\bm{\Sigma}_{\bm{a}}$ in problem~\eqref{eq:mle} is $LM\times LM,$ which is much larger than that in the case of uncorrelated channels~\cite{haghighatshoar2018improved,fengler2021non,chen2022phase}, i.e., $L\times L.$
Therefore, the complexity of updating and storing the matrix $\bm{\Sigma}_{\bm{a}}^{-1}$ in the CD algorithm is much higher than that in the case of uncorrelated channels.
In addition, in the case of uncorrelated channels, updating a single coordinate of $\bm{a}$ corresponds to a rank-$1$ update of the covariance matrix, which enables a closed-form solution for each coordinate update~\cite{haghighatshoar2018improved,fengler2021non,chen2022phase}.
However, in the correlated case, due to \eqref{eq:def-cov}, the rank-$1$ update property does not hold, making it impossible to obtain a closed-form solution for coordinate updates.
These differences make problem~\eqref{eq:mle} much more challenging to solve.

\section{Efficient CD Algorithms for Device Activity Detection} 
\label{sec:algorithm}

In this section, we focus on designing efficient algorithms for solving the MLE problem~\eqref{eq:mle}.
Notice that problem~\eqref{eq:mle} is nonconvex due to the concavity of $\log|\bm{\Sigma}_{\bm{a}}|.$
We first present the exact CD algorithm for solving the MLE problem, which solves the one-dimensional optimization subproblem exactly.
In contrast, the proposed inexact CD algorithm solves this subproblem approximately, thereby achieving improved numerical stability and reduced computational cost.
Furthermore, we study a special case that allows us to further reduce the complexity of both the exact CD and inexact CD algorithms by utilizing the low-rank property of the near-field channel correlation matrix.

We first introduce several key symbols and present the mathematical formulations used to solve the MLE problem.
Let $r_n = \operatorname{rank}(\bm{R}_n).$
Then there exists a matrix $\bm{R}^{\frac{1}{2}}_n \in \mathbb{C}^{M \times r_n}$ such that
\begin{equation}
	\bm{R}_n = \bm{R}_n^{\frac{1}{2}} \big(\bm{R}_n^{\frac{1}{2}}\big)^H.
\end{equation}
Let $\bm{X}_n = \bm{R}^{\frac{1}{2}}_n \otimes \bm{s}_{n} \in \mathbb{C}^{LM\times r_n}.$ Then we have
\begin{equation}\label{eq:xx}
	\bm{X}_n \bm{X}_n^H = \bm{R}_n \otimes \left(\bm{s}_{n} \bm{s}_{n}^H\right).
\end{equation}
Let $f(\bm{a})$ denote the objective function of problem~\eqref{eq:mle}. The $n$-th component of the gradient of $f(\bm{a})$ is given by 
\begin{multline}\label{eq:gradient}
	[\nabla f(\bm{a})]_n = \operatorname{tr} \left( \bm{X}_n^H \bm{\Sigma}_{\bm{a}}^{-1} \bm{X}_n \right) - \left\|\bm{X}_n^H \bm{\Sigma}_{\bm{a}}^{-1}  \left(\bm{y}-\overline{\bm{y}}_{\bm{a}}\right) \right\|_2^2 
	\\- 2\operatorname{Re} \left(   \left(\bm{y}-\overline{\bm{y}}_{\bm{a}}\right)^H \bm{\Sigma}_{\bm{a}}^{-1} \left( \overline{\bm{h}}_{n} \otimes \bm{s}_{n} \right) \right).
\end{multline}
Please refer to Appendix~\ref{sec:proof-objective-decrease} for the derivation of \eqref{eq:gradient}.
Next, we present the first-order optimality condition of problem~\eqref{eq:mle}:
\begin{align}\label{eq:kkt}
	\left[\nabla f(\bm{a})\right]_{n}
	\begin{cases}
		\geq 0, & \text{if}~a_{n} = 0;   \\
		\leq 0, & \text{if}~a_{n} = 1; \\
		= 0, &  \text{if}~ 0 < a_{n} < 1,
	\end{cases}\ \forall~n=1,2,\ldots,N,
\end{align}
which is a necessary condition for the optimality of $\bm{a}.$
To quantify the degree to which each coordinate of $\bm{a}$ violates condition~\eqref{eq:kkt}, we define a nonnegative vector as follows: $\mathscr{abc}$
\begin{equation}\label{eq:def-va}
	\V(\bm{a}) \triangleq \left | \operatorname{Proj}(\bm{a} - \nabla f(\bm{a})) - \bm{a} \right | \in \mathbb{R}_{+}^{N},
\end{equation}
where $\operatorname{Proj}(\cdot)$ is the (coordinate-wise) projection operator onto the interval $[0, 1],$ and $|\cdot|$ is the (coordinate-wise) absolute value.
Notice that the first-order optimality condition~\eqref{eq:kkt} is equivalent to $\V(\bm{a}) = \bm{0}.$
Our goal is to find a feasible point $\bm{a}$ such that $\|\V(\bm{a})\| \le \epsilon,$ where $\epsilon>0$ is a given error tolerance.

\subsection{Exact and Inexact CD Algorithms}

We use the framework of randomly permuted CD~\cite{chen2019covariance}, which is one of the most efficient variants of the CD algorithm.
At each iteration, the algorithm randomly permutes the indices of all coordinates and updates them sequentially according to this permutation.
Let $\bm{e}_n\in \mathbb{R}^N$ denote the vector whose $n$-th component is $1$ and the other components are $0.$
For any given coordinate $n,$ the algorithm updates $a_{n}$ by solving the following one-dimensional optimization subproblem:
\begin{multline}\label{eq:one-dim}
	\underset{d \in [-a_{n},\,1-a_{n}]}{\operatorname{minimize}} \quad f(\bm{a} + d\,\bm{e}_n) = \log\left|\bm{\Sigma}_{\bm{a}} + d \, \bm{X}_n \bm{X}_n^H \right|  \\
	+ \left(\bm{y}-\overline{\bm{y}}_{\bm{a}}-d \, \overline{\bm{h}}_{n} \otimes \bm{s}_{n}\right)^H \left(\bm{\Sigma}_{\bm{a}} + d \, \bm{X}_n \bm{X}_n^H \right)^{-1} \\
	\times \left(\bm{y}-\overline{\bm{y}}_{\bm{a}}-d \, \overline{\bm{h}}_{n} \otimes \bm{s}_{n}\right).
\end{multline}
Notice that problem~\eqref{eq:one-dim} generally does not have a closed-form solution, which sharply contrasts with the CD algorithm in the case where the channels are uncorrelated~\cite{haghighatshoar2018improved,fengler2021non,chen2022phase}.

The key to the CD algorithm is the efficient solution of problem~\eqref{eq:one-dim}.
Next we analyze the property of its objective function.
According to the properties of the determinant, we have
\begin{align}\label{eq:det}
	& \hspace{11.5pt} \log\left|\bm{\Sigma}_{\bm{a}} + d \, \bm{X}_n \bm{X}_n^H \right| \nonumber \\
	& = \log \left| \bm{I}_{LM} + d\,\bm{X}_n \bm{X}_n^H \bm{\Sigma}_{\bm{a}}^{-1} \right| + \log \left|\bm{\Sigma}_{\bm{a}}\right| \nonumber \\
	& = \log \left| \bm{I}_{r_n} + d\, \bm{X}_n^H \bm{\Sigma}_{\bm{a}}^{-1} \bm{X}_n \right| + \log \left|\bm{\Sigma}_{\bm{a}}\right|.
\end{align}
Furthermore, applying the Sherman-Morrison-Woodbury formula, we obtain
\begin{multline}\label{eq:inv}
	\left(\bm{\Sigma}_{\bm{a}} + d \, \bm{X}_n \bm{X}_n^H\right)^{-1} = \bm{\Sigma}_{\bm{a}}^{-1} \\ - d\, \bm{\Sigma}_{\bm{a}}^{-1} \bm{X}_n  \left(\bm{I}_{r_n} + d\, \bm{X}_n^H \bm{\Sigma}_{\bm{a}}^{-1} \bm{X}_n\right)^{-1} \bm{X}_n^H \bm{\Sigma}_{\bm{a}}^{-1}.
\end{multline}
Substituting \eqref{eq:det} and \eqref{eq:inv} into \eqref{eq:one-dim}, we can rewrite problem~\eqref{eq:one-dim} as \eqref{eq:one-dim-r}, shown at the top of the next page.
\begin{figure*}
	\begin{subequations}\label{eq:one-dim-r}
		\begin{alignat}{2}
			\underset{d \in [-a_{n},\,1-a_{n}]}{\operatorname{minimize}} & \quad f(\bm{a} + d\,\bm{e}_n) = f(\bm{a})
			+ \log  \left|\bm{I}_{r_n} + d\, \bm{X}_n^H \bm{\Sigma}_{\bm{a}}^{-1} \bm{X}_n \right| \label{eq:one-dim-r-a} \\
			& - 2\, d \operatorname{Re} \left(   \left(\bm{y}-\overline{\bm{y}}_{\bm{a}}\right)^H \bm{\Sigma}_{\bm{a}}^{-1} \left( \overline{\bm{h}}_{n} \otimes \bm{s}_{n} \right) \right)
			+ d^2 \left( \overline{\bm{h}}_{n} \otimes \bm{s}_{n} \right)^H \bm{\Sigma}_{\bm{a}}^{-1} \left( \overline{\bm{h}}_{n} \otimes \bm{s}_{n} \right)  \label{eq:one-dim-r-b}  \\
			& - d \left(\bm{y}-\overline{\bm{y}}_{\bm{a}}\right)^H	\bm{\Sigma}_{\bm{a}}^{-1} \bm{X}_n  \left(\bm{I}_{r_n} + d\, \bm{X}_n^H \bm{\Sigma}_{\bm{a}}^{-1} \bm{X}_n\right)^{-1} \bm{X}_n^H \bm{\Sigma}_{\bm{a}}^{-1}  \left(\bm{y}-\overline{\bm{y}}_{\bm{a}}\right) \label{eq:one-dim-r-c}  \\
			& + 2\, d^2  \operatorname{Re} \left( \left(\bm{y}-\overline{\bm{y}}_{\bm{a}}\right)^H
			\bm{\Sigma}_{\bm{a}}^{-1} \bm{X}_n  \left(\bm{I}_{r_n} + d\, \bm{X}_n^H \bm{\Sigma}_{\bm{a}}^{-1} \bm{X}_n\right)^{-1} \bm{X}_n^H \bm{\Sigma}_{\bm{a}}^{-1} \left( \overline{\bm{h}}_{n} \otimes \bm{s}_{n} \right) \right) \\
			& - d^3 \left( \overline{\bm{h}}_{n} \otimes \bm{s}_{n} \right)^H	\bm{\Sigma}_{\bm{a}}^{-1} \bm{X}_n  \left(\bm{I}_{r_n} + d\, \bm{X}_n^H \bm{\Sigma}_{\bm{a}}^{-1} \bm{X}_n\right)^{-1} \bm{X}_n^H \bm{\Sigma}_{\bm{a}}^{-1}  \left( \overline{\bm{h}}_{n} \otimes \bm{s}_{n} \right). \label{eq:one-dim-r-e} 
		\end{alignat}
	\end{subequations}
	\hrulefill
\end{figure*}
Problem~\eqref{eq:one-dim-r} is computationally more tractable than problem~\eqref{eq:one-dim} for the following reason.
The key to solving problem~\eqref{eq:one-dim-r} is analyzing the determinant and inverse of the low-dimensional matrix $\bm{I}_{r_n} + d\, \bm{X}_n^H \bm{\Sigma}_{\bm{a}}^{-1} \bm{X}_n \in \mathbb{C}^{r_n \times r_n}.$
In contrast, directly solving problem~\eqref{eq:one-dim} involves handling the high-dimensional matrix $\bm{\Sigma}_{\bm{a}} + d \, \bm{R}_n \otimes \left(\bm{s}_{n} \bm{s}_{n}^H\right) \in \mathbb{C}^{LM \times LM},$ where $r_n \le M \le LM.$

We present two methods for solving problem~\eqref{eq:one-dim-r} below.
Once it is solved, we update variables $a_n,$ $\overline{\bm{y}}_{\bm{a}},$ and $\bm{\Sigma}_{\bm{a}}^{-1},$ with $\bm{\Sigma}_{\bm{a}}^{-1}$ being updated according to~\eqref{eq:inv}.
The exact CD and inexact CD algorithms are summarized in Algorithm~\ref{alg:cd}.

\begin{algorithm}[t]
	\caption{Exact and Inexact CD Algorithms for Solving the MLE Problem~\eqref{eq:mle}}
	\label{alg:cd}
	\begin{algorithmic}[1]
		\STATE Initialize $\bm{a} = \bm{0},$ $\boldsymbol\Sigma_{\bm{a}}^{-1} = \sigma_w^{-2}\bm{I}_{LM},$ and $\overline{\bm{y}}_{\bm{a}} = \bm{0};$ 
		\REPEAT
		\STATE Randomly select a permutation $\{ i_1, i_2, \ldots, i_{N} \}$ of the coordinate indices $\{1, 2, \ldots, N\}$ of $\bm{a};$
		\FOR{$n = i_1,$ $i_2, \ldots, i_{N}$}
		\IF{exact CD} \label{alg:cd-if}
		\STATE Apply the eigenvalue decomposition of the matrix $\bm{X}_n^H \bm{\Sigma}_{\bm{a}}^{-1} \bm{X}_n$ to calculate the coefficients of $p_{\mathrm{num}}(d)$ and $p_{\mathrm{den}}(d)$ in \eqref{eq:one-dim-poly}; \label{alg:cd-exact-1} 
		\STATE Apply the root-finding algorithm~\cite{mcnamee2007numerical} to solve problem~\eqref{eq:one-dim-poly} to obtain $\hat{d},$ and set $d = \hat{d};$\label{alg:cd-exact-2}
		\ELSIF{inexact CD}
		\STATE Calculate the coefficients of $p_{\mathrm{approx}}(d)$ in \eqref{eq:p-approx} and select a $\mu \ge 0$ in \eqref{eq:one-dim-approx}; \label{alg:cd-inexact-1}
		\STATE Apply the cubic formula to solve problem~\eqref{eq:one-dim-approx} to obtain $\bar{d},$ and set $d = \bar{d};$ \label{alg:cd-inexact-2}
		\ENDIF \label{alg:cd-endif}
		\STATE $a_{n} \leftarrow a_{n} + d;$
		\STATE $\overline{\bm{y}}_{\bm{a}} \leftarrow \overline{\bm{y}}_{\bm{a}} + d \, \overline{\bm{h}}_{n} \otimes \bm{s}_{n};$
		\STATE $\bm{\Sigma}_{\bm{a}}^{-1} \leftarrow  \bm{\Sigma}_{\bm{a}}^{-1}$ \\
		\hfill $- d\, \bm{\Sigma}_{\bm{a}}^{-1} \bm{X}_n  \left(\bm{I}_{r_n} + d\, \bm{X}_n^H \bm{\Sigma}_{\bm{a}}^{-1} \bm{X}_n\right)^{-1} \bm{X}_n^H \bm{\Sigma}_{\bm{a}}^{-1};$
		\label{alg:cd-update-sigma} 
		\ENDFOR
		\UNTIL $\|\V(\bm{a})\| \le \epsilon;$
		\STATE Output $\bm{a}.$
	\end{algorithmic}
\end{algorithm}

\subsubsection{Solving Problem~\eqref{eq:one-dim-r} Exactly}

In this part, we present a method for solving problem~\eqref{eq:one-dim-r} exactly, which performs well  when $ r_n $ is small.
The main computational complexities are matrix eigenvalue decomposition and polynomial root-finding.

We now reformulate problem~\eqref{eq:one-dim-r} by addressing each term separately.
First, we consider the term $\log|\cdot|$ in \eqref{eq:one-dim-r-a}.
According to the eigenvalue decomposition of the matrix $\bm{X}_n^H \bm{\Sigma}_{\bm{a}}^{-1} \bm{X}_n,$ there exists a unitary matrix $\bm{V} \in \mathbb{C}^{r_n \times r_n}$ and a diagonal matrix $\bm{\Lambda} = \mathrm{diag}\left( \lambda_1, \lambda_2, \ldots, \lambda_{r_n} \right)$ that satisfy
\begin{equation}\label{eq:eig}
	\bm{X}_n^H \bm{\Sigma}_{\bm{a}}^{-1} \bm{X}_n = \bm{V} \bm{\Lambda} \bm{V}^H.
\end{equation}
Then, we have
\begin{equation}\label{eq:poly-det}
	\left|\bm{I}_{r_n} + d\, \bm{X}_n^H \bm{\Sigma}_{\bm{a}}^{-1} \bm{X}_n \right| = \left|\bm{I}_{r_n} + d\, \bm{\Lambda} \right|  = \prod_{i=1}^{r_n} (1 + d\,\lambda_i),
\end{equation}
which is a polynomial of degree $r_n$ in $d.$
Second, the term in \eqref{eq:one-dim-r-b} is a tractable quadratic polynomial.
Next, we consider the terms in \eqref{eq:one-dim-r-c}--\eqref{eq:one-dim-r-e}.
These terms have the same structure: $d^i \cdot \bm{\xi}^H \left(\bm{I}_{r_n} + d\, \bm{X}_n^H \bm{\Sigma}_{\bm{a}}^{-1} \bm{X}_n\right)^{-1} \bm{\eta}$ with $\bm{\xi}, \bm{\eta} \in \mathbb{C}^{r_n}.$ For example, in \eqref{eq:one-dim-r-e}, $i = 3$ and $\bm{\xi} = \bm{\eta} = \bm{X}_n^H \bm{\Sigma}_{\bm{a}}^{-1}  \left( \overline{\bm{h}}_{n} \otimes \bm{s}_{n} \right).$
We next analyze $\bm{\xi}^H \left(\bm{I}_{r_n} + d\, \bm{X}_n^H \bm{\Sigma}_{\bm{a}}^{-1} \bm{X}_n\right)^{-1} \bm{\eta}.$
Applying~\eqref{eq:eig} and setting $\bar{\bm{\xi}} = \bm{V}^H \bm{\xi}$ and $\bar{\bm{\eta}} = \bm{V}^H \bm{\eta},$ we get
\begin{alignat}{2}\label{eq:poly-inv}
	& \hspace{13.3pt} \bm{\xi}^H \left(\bm{I}_{r_n} + d\, \bm{X}_n^H \bm{\Sigma}_{\bm{a}}^{-1} \bm{X}_n\right)^{-1} \bm{\eta} \nonumber \\
	& = \bm{\xi}^H
	\bm{V} \left(\bm{I}_{r_n} + d\, \bm{\Lambda} \right)^{-1} \bm{V}^H \bm{\eta} \nonumber \\
	& = \bar{\bm{\xi}}^H \left(\bm{I}_{r_n} + d\, \bm{\Lambda} \right)^{-1} \bar{\bm{\eta}} \nonumber \\
	& = \sum_{i=1}^{r_n} \frac{\bar{\xi}_i^{*} \, \bar{\eta}_i }{1 + d\,\lambda_i} \nonumber \\
	& = \frac{\sum_{i=1}^{r_n} \left( \bar{\xi}_i^{*} \, \bar{\eta}_i \prod_{k=1,\,k\neq i}^{r_n} (1 + d\,\lambda_k) \right) }{p_{\mathrm{den}}(d)} ,
\end{alignat}
which is a fraction where the denominator $p_{\mathrm{den}}(d) \triangleq \prod_{i=1}^{r_n} (1 + d\,\lambda_i)$ is a polynomial of degree $r_n$ and the numerator is a polynomial of degree $r_n-1.$
Notice that $p_{\mathrm{den}}(d)$ is also the polynomial in \eqref{eq:poly-det}.
By substituting different $\bm{\xi}$ and $\bm{\eta}$ into \eqref{eq:poly-inv}, we can compute the coefficients of the numerator of each term in \eqref{eq:one-dim-r-c}--\eqref{eq:one-dim-r-e}.
We then combine all terms in \eqref{eq:one-dim-r-b}--\eqref{eq:one-dim-r-e} and reduce them to common denominator $p_{\mathrm{den}}(d).$
Finally, we rewrite problem~\eqref{eq:one-dim-r} as
\begin{equation}\label{eq:one-dim-poly}
	\underset{d \in [-a_{n},\,1-a_{n}]}{\operatorname{minimize}} \quad f(\bm{a} + d\,\bm{e}_n) = f(\bm{a}) + \log p_{\mathrm{den}}(d) + \frac{p_{\mathrm{num}}(d)}{p_{\mathrm{den}}(d)},
\end{equation}
where $p_{\mathrm{num}}(d)$ is a polynomial of degree $r_n+2,$ and its coefficients are derived by combining all the terms in \eqref{eq:one-dim-r-b}--\eqref{eq:one-dim-r-e}.

The process for solving problem~\eqref{eq:one-dim-poly} is as follows.
First, we find $d \in [-a_{n},\,1-a_{n}]$ such that the derivative of the objective function in \eqref{eq:one-dim-poly} is zero.
This is equivalent to setting the numerator of the derivative to zero, which gives:
\begin{equation}\label{eq:root-finding}
	p_{\mathrm{den}}(d)\,p_{\mathrm{den}}^{\prime}(d) + p_{\mathrm{den}}(d)\,p_{\mathrm{num}}^{\prime}(d) - p_{\mathrm{den}}^{\prime}(d)\,p_{\mathrm{num}}(d) = 0,
\end{equation}
where $p_{\mathrm{den}}^{\prime}(d)$ and $p_{\mathrm{num}}^{\prime}(d)$ are the derivatives of $p_{\mathrm{den}}(d)$ and $p_{\mathrm{num}}(d),$ respectively.
Notice that the left-hand side in~\eqref{eq:root-finding} is a polynomial of degree $2 \, r_n+1,$ so problem~\eqref{eq:root-finding} can be solved using a polynomial root-finding algorithm~\cite[Chap.~6]{mcnamee2007numerical}.
Next, we choose the solution with the smallest objective function value in~\eqref{eq:one-dim-poly} among the roots of~\eqref{eq:root-finding} and the two boundary points $-a_{n}$ and $1-a_{n}.$
This yields the exact solution to problem~\eqref{eq:one-dim-poly} (i.e., problem~\eqref{eq:one-dim-r}).
The procedure of solving problem~\eqref{eq:one-dim-r} exactly is outlined in lines~\ref{alg:cd-exact-1} and \ref{alg:cd-exact-2} of Algorithm~\ref{alg:cd}.

The above method is only suitable to solve problem~\eqref{eq:one-dim-r} for small $r_n.$
As observed in Section~\ref{sec:compare-alg}, the exact CD algorithm may diverge when $r_n$ is large.
The observed divergence can be attributed to two main factors.
First, for large $r_n,$ the structures of $ p_{\mathrm{den}}(d) $ and $ p_{\mathrm{num}}(d) $ may cause their values to become very close to zero during certain coordinate updates.
As a result, even small numerical errors in computing $ p_{\mathrm{den}}(d) $ and $ p_{\mathrm{num}}(d) $ can lead to significant inaccuracies in the objective function value of problem~\eqref{eq:one-dim-poly}.
Second, the root-finding process in problem~\eqref{eq:root-finding} can produce numerically inaccurate solutions when the leading coefficient of the polynomial becomes excessively large, a scenario that becomes increasingly likely as $r_n$ increases.
To address these issues, we propose a method for solving problem~\eqref{eq:one-dim-r} inexactly, as detailed below.

\subsubsection{Solving Problem~\eqref{eq:one-dim-r} Inexactly}
\label{subsubsec:inexact}

The key idea for solving problem~\eqref{eq:one-dim-r} inexactly is to use the Taylor expansion to approximate the objective function in \eqref{eq:one-dim-r}, thereby avoiding the eigenvalue decomposition and polynomial root-finding.
Consequently, this method performs well when $r_n$ is large.

As follows, we approximate each term in \eqref{eq:one-dim-r}.
First, using \eqref{eq:poly-det} and the Taylor expansion, we have
\begin{align}\label{eq:approx-det}
	\log \left|\bm{I}_{r_n} + d\, \bm{X}_n^H \bm{\Sigma}_{\bm{a}}^{-1} \bm{X}_n \right| & = \sum_{i=1}^{r_n} \log (1 + d\,\lambda_i) \nonumber \\
	& = \sum_{i=1}^{r_n} d\,\lambda_i  + o(d) \nonumber \\ 
	& = d\operatorname{tr} \left( \bm{X}_n^H \bm{\Sigma}_{\bm{a}}^{-1} \bm{X}_n \right) + o(d),
\end{align}
where the little-$o$ notation $o(d)$ satisfies the condition that $o(d)/d \to 0$ as $d \to 0.$
Further, we apply the Taylor expansion to obtain
\begin{align}\label{eq:approx-inv}
	\left(\bm{I} + d\, \bm{X}_n^H \bm{\Sigma}_{\bm{a}}^{-1} \bm{X}_n \right)^{-1} & = \sum_{k=0}^{\infty} \left(-d\, \bm{X}_n^H \bm{\Sigma}_{\bm{a}}^{-1} \bm{X}_n \right)^{k} \nonumber \\
	& = \bm{I}_{r_n} - d\, \bm{X}_n^H \bm{\Sigma}_{\bm{a}}^{-1} \bm{X}_n + o(d).
\end{align}
Substituting \eqref{eq:approx-det} and \eqref{eq:approx-inv} into \eqref{eq:one-dim-r} and ignoring $o(d),$ we obtain an approximation of the objective function in \eqref{eq:one-dim-r}, shown in \eqref{eq:p-approx} at the top of the next page,
\begin{figure*}
	\begin{align}\label{eq:p-approx}
		p_{\mathrm{approx}}(d) & = f(\bm{a}) + d\,\operatorname{tr} \left( \bm{X}_n^H \bm{\Sigma}_{\bm{a}}^{-1} \bm{X}_n \right)  \nonumber \\ 
		& - 2\,d \operatorname{Re} \left(   \left(\bm{y}-\overline{\bm{y}}_{\bm{a}}\right)^H \bm{\Sigma}_{\bm{a}}^{-1} \left( \overline{\bm{h}}_{n} \otimes \bm{s}_{n} \right) \right)
		+ d^2 \left( \overline{\bm{h}}_{n} \otimes \bm{s}_{n} \right)^H \bm{\Sigma}_{\bm{a}}^{-1} \left( \overline{\bm{h}}_{n} \otimes \bm{s}_{n} \right) \nonumber \\
		& - d \left(\bm{y}-\overline{\bm{y}}_{\bm{a}}\right)^H	\bm{\Sigma}_{\bm{a}}^{-1} \bm{X}_n  \left(\bm{I}_{r_n} - d\, \bm{X}_n^H \bm{\Sigma}_{\bm{a}}^{-1} \bm{X}_n\right) \bm{X}_n^H \bm{\Sigma}_{\bm{a}}^{-1}  \left(\bm{y}-\overline{\bm{y}}_{\bm{a}}\right) \nonumber \\
		& + 2\,d^2 \operatorname{Re} \left( \left(\bm{y}-\overline{\bm{y}}_{\bm{a}}\right)^H
		\bm{\Sigma}_{\bm{a}}^{-1} \bm{X}_n  \left(\bm{I}_{r_n} - d\, \bm{X}_n^H \bm{\Sigma}_{\bm{a}}^{-1} \bm{X}_n\right) \bm{X}_n^H \bm{\Sigma}_{\bm{a}}^{-1} \left( \overline{\bm{h}}_{n} \otimes \bm{s}_{n} \right) \right) \nonumber \\
		& - d^3 \left( \overline{\bm{h}}_{n} \otimes \bm{s}_{n} \right)^H	\bm{\Sigma}_{\bm{a}}^{-1} \bm{X}_n  \left(\bm{I}_{r_n} - d\, \bm{X}_n^H \bm{\Sigma}_{\bm{a}}^{-1} \bm{X}_n\right) \bm{X}_n^H \bm{\Sigma}_{\bm{a}}^{-1}  \left( \overline{\bm{h}}_{n} \otimes \bm{s}_{n} \right).
	\end{align}
	\hrulefill
\end{figure*}
which is a quartic polynomial.
Notice that calculating the coefficients in \eqref{eq:p-approx} does not involve the eigenvalue decomposition.
Next, we solve the following problem:
\begin{equation}\label{eq:one-dim-approx}
	\underset{d \in [-a_{n},\,1-a_{n}]}{\operatorname{minimize}} \quad p_{\mathrm{approx}}(d) + \frac{\mu}{2} d^2,
\end{equation}
where $\mu \ge 0$ is a properly selected parameter.
The complexity of solving problem~\eqref{eq:one-dim-approx} is $\mathcal{O}(1)$ because all roots of the derivative of \eqref{eq:one-dim-approx} can be obtained using the cubic formula.
The optimal solution to \eqref{eq:one-dim-approx} can be obtained by selecting the one with the smallest objective value among the roots and the two boundary points $-a_n$ and $1 - a_{n}.$
The process of solving problem~\eqref{eq:one-dim-r} inexactly is summarized in lines~\ref{alg:cd-inexact-1} and \ref{alg:cd-inexact-2} of Algorithm~\ref{alg:cd}.

The following theorem guarantees that, for an appropriate $\mu,$ the solution to problem~\eqref{eq:one-dim-approx} sufficiently decreases the objective function of problem~\eqref{eq:one-dim-r}.
\begin{theorem}\label{theorem:objective-decrease}
	Let $\lip > 0$ denote the Lipschitz constant of $\nabla f(\bm{a}).$
	There exists a constant $c>0$ such that for $\mu \ge \lip + c,$ the solution $\bar{d}$ to problem~\eqref{eq:one-dim-approx} satisfies
	\begin{equation}\label{eq:objective-decrease}
		f(\bm{a} + \bar{d}\,\bm{e}_n) \le f(\bm{a}) - \frac{[ \V(\bm{a}) ]_n^2}{2(\mu + c)},
	\end{equation}
	where $\V(\bm{a})$ is defined in \eqref{eq:def-va}.
\end{theorem}
\begin{IEEEproof}
	Please see Appendix~\ref{sec:proof-objective-decrease}. 
\end{IEEEproof}

Theorem~\ref{theorem:objective-decrease} demonstrates that, at each coordinate update, although the inexact CD algorithm does not solve problem~\eqref{eq:one-dim-r} exactly, it finds a good approximate solution that sufficiently decreases the objective function, especially when $[ \V(\bm{a}) ]_n$ is large.
Furthermore, building upon the techniques introduced in \cite{wright2015coordinate}, this theorem also establishes the convergence of the inexact CD algorithm to a stationary point.

\subsubsection{Complexity}

In Algorithm~\ref{alg:cd}, the dominant computational complexity of updating a coordinate $a_n$ includes: 
(i) solving the one-dimensional problem~\eqref{eq:one-dim-r}, and (ii) updating matrix $\bm{\Sigma}_{\bm{a}}^{-1}$ with complexity $\mathcal{O}((LM)^2r_n + r_n^3).$
The complexity of solving problem~\eqref{eq:one-dim-r} exactly is $\mathcal{O}(r_n^3),$ which includes the eigenvalue decomposition of matrix $\bm{X}_n^H \bm{\Sigma}_{\bm{a}}^{-1} \bm{X}_n$ with complexity $\mathcal{O}(r_n^3)$ and the root-finding of the polynomial of degree $2 \, r_n+1$ with complexity $\mathcal{O}(r_n^3).$
The complexity of solving problem~\eqref{eq:one-dim-r} inexactly is $\mathcal{O}(1).$
Therefore, the overall complexity of updating $a_n$ is $\mathcal{O}((LM)^2r_n + r_n^3).$

In massive MIMO systems, updating and storing $\bm{\Sigma}_{\bm{a}}^{-1} \in \mathbb{C}^{LM \times LM}$ dominate the complexity of both the exact and inexact CD algorithms.
In the next subsection, we consider a special case where the complexity of updating and storing $\bm{\Sigma}_{\bm{a}}^{-1}$ can be significantly reduced.

\subsubsection{Differences between Presented Algorithms and Those in~\cite{wang2023covariance}}

First, we emphasize that the optimization problem~\eqref{eq:mle} is significantly different from problem (7) in~\cite{wang2023covariance}.
This difference is primarily due to two key factors:
(i) the dimension of the covariance matrix $\bm{\Sigma}_{\bm{a}}$ is significantly larger than that in~\cite{wang2023covariance}; and
(ii) the rank-1 update property, which holds for the covariance matrices in~\cite{wang2023covariance}, does not apply to $\bm{\Sigma}_{\bm{a}}.$

Next, we compare the exact CD algorithm with the CD algorithm in~\cite{wang2023covariance}. Both of them transform the optimization problem into a polynomial root-finding problem, but unlike~\cite{wang2023covariance} where coefficients are computed analytically, this paper proposes applying eigenvalue decomposition to determine the polynomial coefficients.

Finally, we compare the inexact CD algorithm with the inexact CD algorithm in~\cite{wang2023covariance}.
The main difference between the two algorithms is in their handling of the parameter $\mu.$
In~\cite{wang2023covariance}, the parameter $\mu$ is dynamically adjusted at each coordinate update.
However, in this paper, the parameter $\mu$ is fixed to a sufficiently large positive number to ensure convergence.
This difference arises because in~\cite{wang2023covariance}, the computational complexity of the objective function is $\mathcal{O}(1),$ implying that adjusting $\mu$ does not incur any additional computational cost.
In contrast, in this paper, adjusting $\mu$ would result in a computational complexity for the objective function of at least $\mathcal{O}((LM)^2r_n + r_n^3).$
Consequently, we fix $\mu$ to reduce the complexity of coordinate updates.

\subsection{A Special Case Study}

In this subsection, we consider a special case in which the presented exact CD and inexact CD can be accelerated.
The motivation arises from the fact that within a cell, there may be only a few devices in the near-field region of the BS, whereas the majority of devices are located in the far-field region.
Furthermore, we assume that for each device in the near-field region, the number of scatterers $\bar{\ell}_n$ is small.
In this scenario, we can model the near-field channel as correlated channel with low-rank correlation matrix (since $\mathrm{rank}(\bm{R}_n) \le \bar{\ell}_n$), and assume that the far-field channel is uncorrelated.
In this scenario, we only focus on reducing the complexity of updating and storing $\bm{\Sigma}_{\bm{a}}^{-1}$ by taking advantage of the low-rank properties of the near-field correlation matrices. Below, we formally present the assumption for this specific case.
\begin{assumption}\label{assu:rank-deficient}
	Assume that there exists $N_{\mathrm{corr}} \le N$ such that for $1 \le n \le N_{\mathrm{corr}},$ the correlation matrix $\bm{R}_n$ of device $n$ is a low-rank positive semidefinite matrix, and for $N_{\mathrm{corr}}+1 \le n \le N,$ the correlation matrix is $\bm{R}_n = g_n \bm{I}_M,$ where $g_n > 0$ is the large-scale fading coefficient.
	We further assume that the summation of the correlation matrices of the first $N_{\mathrm{corr}}$ devices is rank-deficient, i.e.,
	\begin{equation}\label{eq:sum-rank-deficient}
		r^{\prime} \triangleq \mathrm{rank}\left(\sum_{n=1}^{N_{\mathrm{corr}}}\bm{R}_n\right) < M.
	\end{equation}
\end{assumption}

Notice that $r^{\prime} \le \sum_{n=1}^{N_{\mathrm{corr}}} r_n.$
Then, the limitation to Assumption~\ref{assu:rank-deficient} is twofold: first, the number of devices whose channels are correlated needs to be sufficiently smaller than the number of antennas; second, their channel correlation matrices must be low-rank.
For example, consider a scenario where the number of antennas at the BS is $M = 128.$
There are $N = 200$ potential devices in the cell, but only $N_{\mathrm{corr}} = 20$ devices are in the near-field region, and the channels of the other $180$ devices are uncorrelated.
Assuming the number of scatterers in the near-field channel is $\bar{\ell}_n = 4$ (i.e., $r_n \le 4$), we then have $r^{\prime} \le 20 \times 4 = 80 < 128 = M.$

The following proposition demonstrates that under Assumption~\ref{assu:rank-deficient}, $\bm{R}_n$ ($1\le n \le N_{\mathrm{corr}}$) can be transformed into a matrix where only the upper-left $r^{\prime} \times r^{\prime}$ submatrix is nonzero.
\begin{proposition}\label{prop:submatrix}
	Under Assumption~\ref{assu:rank-deficient}, there exists a unitary matrix $\bm{U} \in \mathbb{C}^{M\times M}$ such that 
	\begin{equation}\label{eq:R-diag}
		\bm{U}^H \bm{R}_n \bm{U} =
		\mathrm{diag}(\widehat{\bm{R}}_n, \bm{0}), \quad  1 \le n \le N_{\mathrm{corr}},
	\end{equation}
	where $\widehat{\bm{R}}_1, \widehat{\bm{R}}_2, \ldots, \widehat{\bm{R}}_{N_{\mathrm{corr}}} \in \mathbb{C}^{r^{\prime} \times r^{\prime}}.$ The matrix $\bm{U}$ can be computed by performing the eigenvalue decomposition on the matrix $\sum_{n=1}^{N_{\mathrm{corr}}}\bm{R}_n,$ i.e., finding a unitary matrix $\bm{U}$ such that
	\begin{equation}\label{eq:eig-R}
		\bm{U}^H \left(\sum_{n=1}^{N_{\mathrm{corr}}}\bm{R}_n\right) \bm{U} = \operatorname{diag}(\lambda_1, \lambda_2, \ldots, \lambda_{r^{\prime}}, 0, \ldots, 0),
	\end{equation}
	where $\lambda_1 \ge \lambda_2 \ge \ldots \ge \lambda_{r^{\prime}} > 0.$
\end{proposition}
\begin{IEEEproof}
	Please see Appendix~\ref{sec:proof-submatrix}. 
\end{IEEEproof}

Furthermore, once the matrix $\bm{U}$ is obtained, we can transform the MLE problem~\eqref{eq:mle} to obtain a more tractable problem as presented in the following theorem.
\begin{theorem}\label{theorem:mle-u}
	The following problem is equivalent to problem~\eqref{eq:mle}:
	\begin{subequations}\label{eq:mle-u}
		\begin{alignat}{2}
			&\underset{\bm{a}}{\operatorname{minimize}}    &\quad& \log|\bm{\Sigma}^{\prime}_{\bm{a}}| +   \left(\bm{y}^{\prime}-\overline{\bm{y}}^{\prime}_{\bm{a}}\right)^H \left(\bm{\Sigma}^{\prime}_{\bm{a}}\right)^{-1} \left(\bm{y}^{\prime}-\overline{\bm{y}}^{\prime}_{\bm{a}}\right)\\
			&\operatorname{subject\,to} &      &a_{n} \in [0,1], \,\forall \, n=1,2,\ldots,N,
		\end{alignat}
	\end{subequations}
	where
	\begin{align}
		\bm{\Sigma}^{\prime}_{\bm{a}} & = \big(\bm{U}^H \otimes \bm{I}_L\big) \bm{\Sigma}_{\bm{a}} \big(\bm{U} \otimes \bm{I}_L\big) \nonumber \\
		& \label{eq:sigma-u} \overset{(a)}{=} \sum_{n=1}^{N_{\mathrm{corr}}} a_{n} \operatorname{diag}(\widehat{\bm{R}}_n, \bm{0}) \otimes \left(\bm{s}_{n} \bm{s}_{n}^H\right) \nonumber \\
		& \qquad + \sum_{n=N_{\mathrm{corr}}+1}^{N} a_{n}\, g_n \bm{I}_M \otimes \left(\bm{s}_{n} \bm{s}_{n}^H\right) + \sigma_w^2 \bm{I}_{LM}, \\
		\label{eq:y-u} \bm{y}^{\prime} & = \left(\bm{U}^H \otimes \bm{I}_L\right) \bm{y}, \\ 
		\label{eq:y-overline-u} \overline{\bm{y}}^{\prime}_{\bm{a}} & = \mathbb{E}[\bm{y}^{\prime}] =  \left(\bm{U}^H \otimes \bm{I}_L\right) \overline{\bm{y}}_{\bm{a}} \overset{(b)}{=} \sum_{n=1}^{N}a_{n} \left(\bm{U}^H \overline{\bm{h}}_{n}\right) \otimes \bm{s}_{n},
	\end{align}
	where the matrix $\bm{U}$ is obtained in Proposition~\ref{prop:submatrix}.
\end{theorem}
\begin{IEEEproof}
	Please see Appendix~\ref{sec:proof-mle-u}. 
\end{IEEEproof}
An intuitive explanation of Theorem~\ref{theorem:mle-u} is as follows.
It is easy to verify that problem~\eqref{eq:mle-u} corresponds to minimizing the negative log-likelihood function of $\bm{y}^{\prime}$ in \eqref{eq:y-u}, i.e., $- \log p(\bm{y}^{\prime}\,|\,\bm{a}).$
According to the property of the transformation of the distribution function, we have
\begin{equation}\label{eq:prob-Jacobi}
	p(\bm{y}^{\prime}\,|\,\bm{a}) = p(\bm{y}\,|\,\bm{a}) \cdot \left| \bm{U}^H \otimes \bm{I}_L \right| = p(\bm{y}\,|\,\bm{a}).
\end{equation}
Therefore, problem~\eqref{eq:mle-u} (i.e., minimizing $- \log p(\bm{y}^{\prime}\,|\,\bm{a})$) is equivalent to problem~\eqref{eq:mle} (i.e., minimizing $- \log p(\bm{y}\,|\,\bm{a})$).

Notice that problem~\eqref{eq:mle-u} has a similar structure to problem~\eqref{eq:mle}.
Therefore, we can apply both exact and inexact CD algorithms in Algorithm~\ref{alg:cd} to solve it.
In line~\ref{alg:cd-update-sigma} of Algorithm~\ref{alg:cd}, we need to update the matrix $(\bm{\Sigma}^{\prime}_{\bm{a}})^{-1}$ with complexity $\mathcal{O}((LM)^2r_n + r_n^3).$
However, $(\bm{\Sigma}^{\prime}_{\bm{a}})^{-1}$ can be updated and stored with significantly lower complexity. Specifically, $\bm{\Sigma}^{\prime}_{\bm{a}}$ in \eqref{eq:sigma-u} is a block diagonal matrix, i.e.,
\begin{equation}\label{eq:sigma-diag}
	\bm{\Sigma}^{\prime}_{\bm{a}} = \mathrm{diag} \big( \widehat{\bm{\Sigma}}^{\prime}_{\bm{a}}, \underbrace{   \widetilde{\bm{\Sigma}}^{\prime}_{\bm{a}}, \widetilde{\bm{\Sigma}}^{\prime}_{\bm{a}}, \ldots, \widetilde{\bm{\Sigma}}^{\prime}_{\bm{a}}   }_{M-r^{\prime}} \big),
\end{equation}
where
\begin{align}
	\widehat{\bm{\Sigma}}^{\prime}_{\bm{a}} & = \sum_{n=1}^{N_{\mathrm{corr}}} a_{n} \widehat{\bm{R}}_n \otimes \left(\bm{s}_{n} \bm{s}_{n}^H\right) \nonumber \\
	& \qquad + \sum_{n=N_{\mathrm{corr}}+1}^{N} a_{n}\, g_n \bm{I}_{r^{\prime}} \otimes \left(\bm{s}_{n} \bm{s}_{n}^H\right) + \sigma_w^2 \bm{I}_{Lr^{\prime}} \nonumber \\
	& \overset{(a)}{=} \sum_{n=1}^{N} a_{n} \widehat{\bm{X}}_n \widehat{\bm{X}}_n^H + \sigma_w^2 \bm{I}_{Lr^{\prime}} \in \mathbb{C}^{Lr^{\prime} \times Lr^{\prime}}, \label{eq:sigma-hat} \\
	\widetilde{\bm{\Sigma}}^{\prime}_{\bm{a}} & = \sum_{n=N_{\mathrm{corr}}+1}^{N} a_{n} g_n \bm{s}_{n} \bm{s}_{n}^H + \sigma_w^2 \bm{I}_{L} \in \mathbb{C}^{L \times L}, \label{eq:sigma-tilde}
\end{align}
and $(a)$ is because we set $\widehat{\bm{X}}_n = \widehat{\bm{R}}^{\frac{1}{2}}_n \otimes \bm{s}_{n} \in \mathbb{C}^{Lr^{\prime}\times r_n}$ for $1 \le n \le N_{\mathrm{corr}},$ where $\widehat{\bm{R}}^{\frac{1}{2}}_n \in \mathbb{C}^{r^{\prime}\times r_n}$ satisfies $\widehat{\bm{R}}^{\frac{1}{2}}_n \big(\widehat{\bm{R}}^{\frac{1}{2}}_n\big)^H = \widehat{\bm{R}}_n,$ and $\widehat{\bm{X}}_n = \sqrt{g_n} \bm{I}_{r^{\prime}} \otimes \bm{s}_{n} \in \mathbb{C}^{Lr^{\prime}\times r^{\prime}}$ for $N_{\mathrm{corr}}+1 \le n \le N.$
Therefore, in a coordinate update, we can obtain the updated $(\bm{\Sigma}^{\prime}_{\bm{a}})^{-1}$ by updating two low-dimensional matrices $\big(\widehat{\bm{\Sigma}}^{\prime}_{\bm{a}}\big)^{-1}$ and $\big(\widetilde{\bm{\Sigma}}^{\prime}_{\bm{a}}\big)^{-1}.$ 
The details are summarized in Algorithm~\ref{alg:update-sigma}.
Notice that the complexity of updating $(\bm{\Sigma}^{\prime}_{\bm{a}})^{-1}$ in Algorithm~\ref{alg:update-sigma} is $\mathcal{O}((Lr^{\prime})^2r_n + r_n^3)$ for $1 \le n \le N_{\mathrm{corr}}$ and $\mathcal{O}((Lr^{\prime})^2r^{\prime} + (r^{\prime})^3)$ for $N_{\mathrm{corr}}+1 \le n \le N,$ which is significantly lower than $\mathcal{O}((LM)^2r_n + r_n^3)$ because $r^{\prime} < M.$
Furthermore, we only need to store $\big(\widehat{\bm{\Sigma}}^{\prime}_{\bm{a}}\big)^{-1}$ and $\big(\widetilde{\bm{\Sigma}}^{\prime}_{\bm{a}}\big)^{-1}$ instead of $(\bm{\Sigma}^{\prime}_{\bm{a}})^{-1},$
because all matrix-vector multiplications involving $(\bm{\Sigma}^{\prime}_{\bm{a}})^{-1}$ can be expressed using $\big(\widehat{\bm{\Sigma}}^{\prime}_{\bm{a}}\big)^{-1}$ and $\big(\widetilde{\bm{\Sigma}}^{\prime}_{\bm{a}}\big)^{-1}.$
Specifically, for any vectors $\bm{\xi}, \bm{\eta} \in \mathbb{C}^{LM},$ we have
\begin{multline}
	\bm{\xi}^H (\bm{\Sigma}^{\prime}_{\bm{a}})^{-1} \bm{\eta} = \bm{\xi}_{1:Lr^{\prime}}^H \big(\widehat{\bm{\Sigma}}^{\prime}_{\bm{a}}\big)^{-1} \bm{\eta}_{1:Lr^{\prime}} \\ + \sum_{m=r^{\prime}+1}^{M} \bm{\xi}_{L(m-1)+1:Lm}^H \big(\widetilde{\bm{\Sigma}}^{\prime}_{\bm{a}}\big)^{-1} \bm{\eta}_{L(m-1)+1:Lm}.
\end{multline}

\begin{algorithm}[t]
	\caption{Updating $(\bm{\Sigma}^{\prime}_{\bm{a}})^{-1}$ When $a_n$ is Increased by $d$ in One Coordinate Update}
	\label{alg:update-sigma}
	\begin{algorithmic}[1]
		\STATE Input $n,$ $d,$ $\big(\widehat{\bm{\Sigma}}^{\prime}_{\bm{a}}\big)^{-1},$ and  $\big(\widetilde{\bm{\Sigma}}^{\prime}_{\bm{a}}\big)^{-1};$
		\IF{$1 \le n \le N_{\mathrm{corr}}$} 
		\STATE $\big(\widehat{\bm{\Sigma}}^{\prime}_{\bm{a}}\big)^{-1} \leftarrow  \big(\widehat{\bm{\Sigma}}^{\prime}_{\bm{a}}\big)^{-1} - d\, \big(\widehat{\bm{\Sigma}}^{\prime}_{\bm{a}}\big)^{-1} \widehat{\bm{X}}_n $
		\\ \qquad \qquad $ \times \big(\bm{I}_{r_n} + d\, \widehat{\bm{X}}_n^H \big(\widehat{\bm{\Sigma}}^{\prime}_{\bm{a}}\big)^{-1} \widehat{\bm{X}}_n\big)^{-1} \widehat{\bm{X}}_n^H \big(\widehat{\bm{\Sigma}}^{\prime}_{\bm{a}}\big)^{-1};$ 
		\ELSIF{$N_{\mathrm{corr}}+1 \le n \le N$} 
		\STATE $\big(\widehat{\bm{\Sigma}}^{\prime}_{\bm{a}}\big)^{-1} \leftarrow  \big(\widehat{\bm{\Sigma}}^{\prime}_{\bm{a}}\big)^{-1} - d\, \big(\widehat{\bm{\Sigma}}^{\prime}_{\bm{a}}\big)^{-1} \widehat{\bm{X}}_n $
		\\ \qquad \qquad $ \times \big(\bm{I}_{r^{\prime}} + d\, \widehat{\bm{X}}_n^H \big(\widehat{\bm{\Sigma}}^{\prime}_{\bm{a}}\big)^{-1} \widehat{\bm{X}}_n\big)^{-1} \widehat{\bm{X}}_n^H \big(\widehat{\bm{\Sigma}}^{\prime}_{\bm{a}}\big)^{-1};$
		\STATE $\big(\widetilde{\bm{\Sigma}}^{\prime}_{\bm{a}}\big)^{-1} \leftarrow \big(\widetilde{\bm{\Sigma}}^{\prime}_{\bm{a}}\big)^{-1} -  d \, g_n \,\big(\widetilde{\bm{\Sigma}}^{\prime}_{\bm{a}}\big)^{-1} \bm{s}_{n} \bm{s}_{n}^H \big(\widetilde{\bm{\Sigma}}^{\prime}_{\bm{a}}\big)^{-1} $
		\\ \qquad \qquad  $ / \big(1\, +\, d\, g_{n}\, \bm{s}_{n}^H \big(\widetilde{\bm{\Sigma}}^{\prime}_{\bm{a}}\big)^{-1} \bm{s}_{n}\big) ;$ 
		\ENDIF 
		\STATE Output $(\bm{\Sigma}^{\prime}_{\bm{a}})^{-1} = \mathrm{diag} \big( \big(\widehat{\bm{\Sigma}}^{\prime}_{\bm{a}}\big)^{-1}, \big(\widetilde{\bm{\Sigma}}^{\prime}_{\bm{a}}\big)^{-1},  \ldots, \big(\widetilde{\bm{\Sigma}}^{\prime}_{\bm{a}}\big)^{-1} \big).$
	\end{algorithmic}
\end{algorithm}

Under Assumption~\ref{assu:rank-deficient}, Algorithm~\ref{alg:update-sigma} reduces the complexity of updating $(\bm{\Sigma}^{\prime}_{\bm{a}})^{-1}$ from $\mathcal{O}((LM)^2r_n + r_n^3)$ to at most $\mathcal{O}((Lr^{\prime})^2r^{\prime} + (r^{\prime})^3),$ which is independent of $M.$ Therefore, it will significantly reduce the complexity of exact and inexact CD algorithms in massive MIMO systems.
The one-dimensional optimization subproblem can be solved either exactly or inexactly using techniques similar to those in Algorithm~\ref{alg:cd}, with the same complexity.
We omit these details here for simplicity.

If the condition \eqref{eq:sum-rank-deficient} in Assumption~\ref{assu:rank-deficient} does not hold, i.e., $\mathrm{rank}\left(\sum_{n=1}^{N_{\mathrm{corr}}}\bm{R}_n\right) = M,$ we can design a low-complexity algorithm inspired by Algorithm~\ref{alg:update-sigma} to obtain an approximate solution to the MLE problem~\eqref{eq:mle}.
We describe it briefly as follows:
First, we perform the eigenvalue decomposition on the matrix $\sum_{n=1}^{N_{\mathrm{corr}}}\bm{R}_n$ to obtain the unitary matrix $\bm{U}.$
Although the matrix $\bm{U}^H \bm{R}_n \bm{U}$ does not retain the property in \eqref{eq:R-diag}, we can select a parameter $r^{\prime\prime}$ such that $r^{\prime\prime} < M$ and set $\widehat{\bm{R}}_n$ to the upper-left $r^{\prime\prime} \times r^{\prime\prime}$ submatrix of $\bm{U}^H \bm{R}_n \bm{U},$ i.e., $\widehat{\bm{R}}_n = \left(\bm{U}^H \bm{R}_n \bm{U}\right)_{1:r^{\prime\prime},\, 1:r^{\prime\prime}}$ for $1 \le n \le N_{\mathrm{corr}}.$
Next, we substitute $\widehat{\bm{R}}_n$ into \eqref{eq:sigma-u} and apply the exact or inexact CD algorithm to solve problem~\eqref{eq:mle-u}, with $(\bm{\Sigma}^{\prime}_{\bm{a}})^{-1}$ updated by Algorithm~\ref{alg:update-sigma}.
The complexity of updating $(\bm{\Sigma}^{\prime}_{\bm{a}})^{-1}$ is then at most $\mathcal{O}((Lr^{\prime\prime})^2r^{\prime\prime} + (r^{\prime\prime})^3).$
For a properly selected $r^{\prime\prime},$ the output of the algorithm is a high-quality approximate solution to problem~\eqref{eq:mle}. This approximate solution can be used as an initial point for Algorithm~\ref{alg:cd} to accelerate the convergence of the algorithm.

\section{Detection Performance Analysis}
\label{sec:analysis}

In this section, we analyze how the channel correlation matrices affect the detection performance of the MLE model~\eqref{eq:mle}.
Our analysis is based on the low-rank properties of near-field spatial correlation matrices~\cite{liu2023near,lu2024tutorial,dong2022near,dong2024characterizing}, focusing on two perspectives: signal statistical dimension and identifiability of the true activity vector.
On the one hand, low-rank correlation matrices may reduce the signal statistical dimension compared to uncorrelated channels, especially when the ratio $N/L$ is small.
This reduction of signal statistical dimension will lead to a loss in detection performance.
On the other hand, covariance identifiability under correlated channels is more likely to hold, implying that detection performance under correlated channels is better than that under uncorrelated channels when the ratio $N/L$ is large.

In order to reveal how the channel correlation affects the detection performance of the MLE model, the analysis in this section focuses on the following two cases, which differ only in the correlation of the channels:
\begin{enumerate}[{Case} I:]
	\item \label{item:corr} For each device $n,$ the channel $\bm{h}_{n} \in \mathbb{C}^M$ follows $\mathcal{CN}\big(\overline{\bm{h}}_n,\bm{R}_n\big),$ where $\overline{\bm{h}}_n$ and $\bm{R}_n$ are defined in \eqref{eq:los-channel} and \eqref{eq:R-structure}.
	\item \label{item:uncorr} For each device $n,$ the channel $\bm{h}_{n} \in \mathbb{C}^M$ follows $\mathcal{CN}\big(\overline{\bm{h}}_n,g_n\bm{I}_M\big),$ where $g_n = \frac{1}{M} \mathrm{tr}(\bm{R}_n)$ is the large-scale fading coefficient.
\end{enumerate}
We assume that the set of signature sequences is the same in both Case~\ref{item:corr} and Case~\ref{item:uncorr}, with each sequence $\bm{s}_{n}$ being randomly generated from a specific distribution, such as the uniform distribution over the discrete set $\qam^L,$ as in~\cite{wang2023covariance}.

\subsection{Signal Statistical Dimension}
\label{subsec:dim}

The main idea of this subsection is to define and compare signal statistical dimensions in Case~\ref{item:corr} and Case~\ref{item:uncorr}.
We infer that a larger signal statistical dimension corresponds to better detection performance because the received signal can be transformed by a unitary transformation.
In the transformed signal vector, only the covariance matrix of a subvector provides information about $\bm{a}.$
The length of this subvector is the signal statistical dimension.
We then compare the upper bounds of the signal statistical dimensions in Case~\ref{item:corr} and Case~\ref{item:uncorr}, which indicates the relationship between the system parameters and the signal statistical dimensions.

We now define the signal statistical dimensions $D_{\mathrm{I}}$ and $D_{\mathrm{II}}$ in Case~\ref{item:corr} and Case~\ref{item:uncorr} respectively:
\begin{align}
	D_{\mathrm{I}} & \triangleq \mathrm{rank}\left( \sum_{n=1}^{N} \bm{R}_n \otimes \left(\bm{s}_{n} \bm{s}_{n}^H\right) \right) \label{eq:def-D-I} ,\\
	D_{\mathrm{II}} & \triangleq \mathrm{rank}\left( \sum_{n=1}^{N} g_n \bm{I}_M \otimes \left(\bm{s}_{n} \bm{s}_{n}^H\right) \right) \label{eq:def-D-II} .
\end{align}
The statistical dimensions $D_{\mathrm{I}}$ and $D_{\mathrm{II}}$ are closely related to the detection performance in Case~\ref{item:corr} and Case~\ref{item:uncorr} due to the following reason.
Without loss of generality, we consider Case~\ref{item:corr}.
Notice that the received signal $\bm{y} \in \mathbb{C}^{LM}$ follows $\mathcal{CN}(\overline{\bm{y}}_{\bm{a}},\bm{\Sigma}_{\bm{a}}),$ where $\overline{\bm{y}}_{\bm{a}}$ and $\bm{\Sigma}_{\bm{a}}$ are given in \eqref{eq:def-y-bar} and \eqref{eq:def-cov}.
Using techniques similar to Proposition~\ref{prop:submatrix}, we have that there exists a unitary matrix $\bm{V} \in \mathbb{C}^{LM \times LM}$ such that 
\begin{equation}
	\bm{V}^H \left(\bm{R}_n \otimes \left(\bm{s}_{n} \bm{s}_{n}^H\right)\right) \bm{V} = 
	\operatorname{diag}(\bm{\Phi}_n , \bm{0}),
\end{equation}
where $\bm{\Phi}_n \in \mathbb{C}^{D_{\mathrm{I}} \times D_{\mathrm{I}}}.$
Defining the transformed received signal $\bm{y}^{\prime} = \bm{V}^H \bm{y},$ we have $\bm{y}^{\prime} \sim \mathcal{CN} ( \overline{\bm{y}}_{\bm{a}}^{\prime}, \bm{\Sigma}_{\bm{a}}^{\prime} ),$ where $\overline{\bm{y}}_{\bm{a}}^{\prime} = \bm{V}^H \overline{\bm{y}}_{\bm{a}},$ and
\begin{equation}\label{eq:def-cov-hat}
	\bm{\Sigma}_{\bm{a}}^{\prime} = \bm{V}^H \bm{\Sigma}_{\bm{a}} \bm{V} = \sum_{n=1}^{N} a_{n} \operatorname{diag}(\bm{\Phi}_n , \bm{0}) + \sigma_w^2 \bm{I}_{LM}.
\end{equation}
We observe from \eqref{eq:def-cov-hat} that $\bm{\Sigma}_{\bm{a}}^{\prime} \in \mathbb{C}^{LM \times LM}$ is a block diagonal matrix and its lower-right submatrix $\big(\bm{\Sigma}_{\bm{a}}^{\prime}\big)_{D_{\mathrm{I}}+1:LM,D_{\mathrm{I}}+1:LM} = \sigma_w^2 \bm{I}_{LM-D_{\mathrm{I}}}$ is independent of $\bm{a}.$
Therefore, in the transformed received signal $\bm{y}^{\prime},$ only the covariance of subvector $\bm{y}_{1:D_{\mathrm{I}}}^{\prime}$ depends on $\bm{a},$ while the covariance of subvector $\bm{y}_{D_{\mathrm{I}}+1:LM}^{\prime}$ is just $\sigma_w^2 \bm{I}_{LM-D_{\mathrm{I}}}.$
Applying the property of the transformation of the distribution function, we can rewrite the log-likelihood function as
\begin{align}\label{eq:log-likehood}
	\log p(\bm{y} \, | \, \bm{a}) & = \log p(\bm{y}^{\prime} \, | \, \bm{a}) \nonumber  \\
	& \overset{(a)}{=} \log p(\bm{y}_{1:D_{\mathrm{I}}}^{\prime} \, | \, \bm{a}) + \log p(\bm{y}_{D_{\mathrm{I}}+1:LM}^{\prime} \, | \, \bm{a}),
\end{align}
where $(a)$ is due to the fact that $\bm{y}_{1:D_{\mathrm{I}}}^{\prime}$ and $\bm{y}_{D_{\mathrm{I}}+1:LM}^{\prime}$ are independent of each other for a given $\bm{a}.$
The log-likelihood function in \eqref{eq:log-likehood} implies that when solving the MLE problem~\eqref{eq:mle}, only the mean and covariance of subvector $\bm{y}_{1:D_{\mathrm{I}}}^{\prime}$ and the mean of subvector $\bm{y}_{D_{\mathrm{I}}+1:LM}^{\prime}$ are used for activity detection.
This is why we refer to $D_{\mathrm{I}}$ as the statistical dimension.
Furthermore, if $\overline{\bm{h}}_n = \bm{0}$ for all $n,$ then subvector $\bm{y}_{D_{\mathrm{I}}+1:LM}^{\prime}$ is simply the noise that follows $\mathcal{CN} (\bm{0}, \sigma_w^2 \bm{I}_{LM-D_{\mathrm{I}}}).$
In this case, only the $D_{\mathrm{I}}$-dimensional vector $\bm{y}_{1:D_{\mathrm{I}}}^{\prime}$ is useful for activity detection.
Therefore, the larger the value of $D_{\mathrm{I}},$ the better the detection performance in Case~\ref{item:corr}.
The same relationship holds for Case~\ref{item:uncorr} with respect to $D_{\mathrm{II}}.$

Since $D_{\mathrm{I}}$ and $D_{\mathrm{II}}$ in \eqref{eq:def-D-I} and \eqref{eq:def-D-II} generally do not have a closed-form expression, we estimate their values by calculating their upper bounds.
In Case~\ref{item:corr}, denote $\bar{r} = \max_{n} \operatorname{rank}\left( \bm{R}_n \right).$ Then we obtain
\begin{align}\label{eq:D-near}
	D_{\mathrm{I}} & \le \sum_{n=1}^{N}  \operatorname{rank}\left( \bm{R}_n \otimes \left(\bm{s}_{n} \bm{s}_{n}^H\right) \right) \nonumber \\
	& = \sum_{n=1}^{N}  \operatorname{rank}\left( \bm{R}_n \right) \operatorname{rank}\left( \bm{s}_{n} \bm{s}_{n}^H\right) \le \bar{r}N.
\end{align}
Since $D_{\mathrm{I}}$ also satisfies $D_{\mathrm{I}} \le LM,$ we have $\bar{D}_{\mathrm{I}} = \min(\bar{r}N, LM)$ as an upper bound on $D_{\mathrm{I}}.$ 
In Case~\ref{item:uncorr}, due to $\operatorname{rank}\left( g_n \bm{I}_M \right) = M,$ and using similar techniques as in \eqref{eq:D-near}, we have the upper bound $\bar{D}_{\mathrm{II}} = \min( MN, LM ) \overset{(a)}{=} LM,$ where $(a)$ is due to the fact that in the random access setting $N \ge L.$
Therefore, we have
\begin{equation}\label{eq:compare-D}
	\bar{D}_{\mathrm{I}} 
	\begin{cases}
		< \bar{D}_{\mathrm{II}} , & \text{if } N/L < M/\bar{r}, \\
		= \bar{D}_{\mathrm{II}} , & \text{if } N/L \ge M/\bar{r}.
	\end{cases}
\end{equation}
The key inference from \eqref{eq:compare-D} is that when the ratio $N/L$ is small, the detection performance in Case~\ref{item:corr} would be worse than that in Case~\ref{item:uncorr}. Whereas when $N/L$ is large, $D_{\mathrm{I}}$ and $D_{\mathrm{II}}$ are close to the same.
Next, we will illustrate from another perspective that the detection performance in Case~\ref{item:corr} is better when $N/L$ is large.

\subsection{Identifiability of True Activity Vector}
\label{subsec:id}

The main idea of this subsection arises from the phase transition analysis in \cite{chen2022phase}, which demonstrates that a crucial condition for the MLE model~\eqref{eq:mle} to perform effectively is that the true activity indicator vector $\bm{a}^{\circ}$ needs to be uniquely identifiable.
This implies that there should be no other vector $\bm{a}^{\prime} \neq \bm{a}^{\circ}$ such that $p(\bm{y} \, | \, \bm{a}^{\prime}) = p(\bm{y} \, | \, \bm{a}^{\circ}).$
We demonstrate that the unique identifiability of $\bm{a}^{\circ}$ is more likely to be achieved in Case~\ref{item:corr} than in Case~\ref{item:uncorr}.
Furthermore, inspired by the unique identifiability of $\bm{a}^{\circ},$ we infer that the mutual interference between different devices is weaker in Case~\ref{item:corr} compared to Case~\ref{item:uncorr}.

We now investigate the identifiability of $\bm{a}^{\circ}$ in both Case~\ref{item:corr} and Case~\ref{item:uncorr}.
Since both $p(\bm{y} \, | \, \bm{a})$ and $p(\bm{y} \, | \, \bm{a}^{\circ})$ are multivariate Gaussian distributions, their equivalence implies that their corresponding means and covariances must be identical.
Therefore, the identifiability of $\bm{a}^{\circ}$ is equivalent to the condition that there is no other $\bm{a}^{\prime} \neq \bm{a}^{\circ}$ such that $\overline{\bm{y}}_{\bm{a}^{\prime}} = \overline{\bm{y}}_{\bm{a}^{\circ}}$ and $\bm{\Sigma}_{\bm{a}^{\prime}} = \bm{\Sigma}_{\bm{a}^{\circ}}.$
Furthermore, we define
\begin{align}
	\mathcal{N}_{\mathrm{I}} & \triangleq \Big\{ \bm{x} \in \mathbb{R}^N  \Big| \textstyle\sum_{n=1}^{N} x_n \bm{R}_n \otimes \left(\bm{s}_{n} \bm{s}_{n}^H\right) = \bm{0}, \nonumber \\
	& \qquad \qquad \qquad \qquad \text{ and } \textstyle\sum_{n=1}^{N} x_n \, \overline{\bm{h}}_n \otimes \bm{s}_n = \bm{0}  \Big\} ,\\
	\mathcal{N}_{\mathrm{II}} & \triangleq \Big\{ \bm{x} \in \mathbb{R}^N  \Big| \textstyle\sum_{n=1}^{N} x_n g_n\bm{I}_M \otimes \left(\bm{s}_{n} \bm{s}_{n}^H\right) = \bm{0}, \nonumber \\
	& \qquad \qquad \qquad \qquad \text{ and } \textstyle\sum_{n=1}^{N} x_n \, \overline{\bm{h}}_n \otimes \bm{s}_n = \bm{0} \Big\} ,\\
	\mathcal{C} & \triangleq \left\{ \left. \bm{x} \in \mathbb{R}^N \,\right| x_n \ge 0 \text{ if } a^{\circ} = 0,\, x_n \le 0 \text{ if } a^{\circ} = 1 \right\}.
\end{align}
The identifiability of $\bm{a}^{\circ}$ in Case~\ref{item:corr} and Case~\ref{item:uncorr} is equivalent to the conditions $\mathcal{N}_{\mathrm{I}} \cap \mathcal{C} = \{ \bm{0} \}$ and $\mathcal{N}_{\mathrm{II}} \cap \mathcal{C} = \{ \bm{0} \},$ respectively.
The following theorem illustrates that $\mathcal{N}_{\mathrm{II}} \cap \mathcal{C} = \{ \bm{0} \}$ is a sufficient condition for $\mathcal{N}_{\mathrm{I}} \cap \mathcal{C} = \{ \bm{0} \}.$

\begin{theorem}\label{theorem:nc}
	If $\mathcal{N}_{\mathrm{II}} \cap \mathcal{C} = \{ \bm{0} \},$ then $\mathcal{N}_{\mathrm{I}} \cap \mathcal{C} = \{ \bm{0} \}$ also holds true.
\end{theorem}
\begin{IEEEproof}
	Please see Appendix~\ref{sec:proof-nc}.
\end{IEEEproof}

Theorem~\ref{theorem:nc} highlights a key observation that the identifiability of $\bm{a}^{\circ}$ is more likely to hold in Case~\ref{item:corr} than in Case~\ref{item:uncorr}.
Furthermore, the scaling law results in \cite{fengler2021non,chen2022phase,wang2023covariance} show that when the system parameters satisfy $K = \mathcal{O}(L^2)$ and under certain methods of generating signature sequences, $\mathcal{N}_{\mathrm{II}} \cap \mathcal{C} = \{ \bm{0} \}$ holds. Therefore, $\mathcal{N}_{\mathrm{I}} \cap \mathcal{C} = \{ \bm{0} \}$ also holds under these conditions, implying that the number of active devices that can be correctly detected by the MLE problem~\eqref{eq:mle} in Case~\ref{item:corr} is at least $K = \mathcal{O}(L^2).$

Notice that if $\overline{\bm{h}}_n = \bm{0}$ for all $n,$ then sets $\mathcal{N}_{\mathrm{I}}$ and $\mathcal{N}_{\mathrm{II}}$ can be represented as
\begin{align}
	\mathcal{N}_{\mathrm{I}} & = \left\{ \bm{x} \in \mathbb{R}^N  \left| \bm{\Psi}^{(I)} \bm{x} = \bm{0} \right. \right\} ,\\
	\mathcal{N}_{\mathrm{II}} & = \left\{ \bm{x} \in \mathbb{R}^N  \left| \bm{\Psi}^{(II)} \bm{x} = \bm{0} \right. \right\} ,
\end{align}
where $\bm{\Psi}^{(I)} \in \mathbb{C}^{(LM)^2\times N}$ and $\bm{\Psi}^{(II)}\in \mathbb{C}^{(LM)^2\times N}$ are matrices whose $n$-th columns are $\bm{\psi}_n^{(I)} = \operatorname{vec}\left(\bm{R}_n \otimes \left(\bm{s}_{n} \bm{s}_{n}^H\right)\right)$ and $\bm{\psi}_n^{(II)} = \operatorname{vec}\left(g_n\bm{I}_M \otimes \left(\bm{s}_{n} \bm{s}_{n}^H\right)\right),$ respectively.
Taking Case~\ref{item:corr} as an example, an heuristic view is that if the columns of $\bm{\Psi}^{(I)}$ are closer to orthogonality, then the condition $\mathcal{N}_{\mathrm{I}} \cap \mathcal{C} = \{ \bm{0} \}$ is more likely to hold, which also implies that the mutual interference between devices is smaller.
To this end, for different devices $n$ and $k,$ we study the following cosine similarity:
\begin{multline}\label{eq:cosine}
	\frac{ \big(\bm{\psi}_n^{(I)}\big)^H  \bm{\psi}_k^{(I)}}{ \| \bm{\psi}_n^{(I)} \|_2 \| \bm{\psi}_k^{(I)} \|_2 } = \frac{\mathrm{tr}(\bm{R}_n \bm{R}_k)}{\| \bm{R}_n \|_F \| \bm{R}_k \|_F} \left( \frac{| \bm{s}_n^H \bm{s}_k|}{ \| \bm{s}_n \|_2 \| \bm{s}_k \|_2} \right)^2 \\
	\overset{(a)}{\le} \left( \frac{| \bm{s}_n^H \bm{s}_k|}{ \| \bm{s}_n \|_2 \| \bm{s}_k \|_2} \right)^2 = \frac{ \big(\bm{\psi}_n^{(II)}\big)^H  \bm{\psi}_k^{(II)}}{ \| \bm{\psi}_n^{(II)} \|_2 \| \bm{\psi}_k^{(II)} \|_2 },
\end{multline}
where $(a)$ follows from the Cauchy-Schwartz inequality $\mathrm{tr}(\bm{R}_n \bm{R}_k) \le \| \bm{R}_n \|_F \| \bm{R}_k \|_F,$ and equality holds if and only if there exists a constant $c$ such that $\bm{R}_n = c\bm{R}_k.$
We observe from \eqref{eq:cosine} that $\bm{\psi}_n^{(I)}$ and $\bm{\psi}_k^{(I)}$ are closer to orthogonal than $\bm{\psi}_n^{(II)}$ and $\bm{\psi}_k^{(II)},$ implying that the mutual interference between devices $n$ and $k$ in Case~\ref{item:corr} is lower than that in Case~\ref{item:uncorr}.
The above analysis demonstrates that the channel correlation among different antennas can reduce the mutual interference between devices and thereby improve the detection performance.
Therefore, when the signal statistical dimensions in Case~\ref{item:corr} and Case~\ref{item:uncorr} are equal (i.e., $N/L$ is large), the MLE model~\eqref{eq:mle} exhibits superior detection performance in Case~\ref{item:corr} (than in Case~\ref{item:uncorr}).

\section{Joint Device Activity and Data Detection}
\label{sec:data-detection}

This section aims to show that the MLE model~\eqref{eq:mle} can be applied to the joint device activity and data detection problem, in which each device is associated with multiple signature sequences and can embed a few information bits~\cite{chen2019covariance}.

Suppose that each active device intends to send $J$ bits of information to the BS, where $J$ is a small integer.
To perform joint activity and data detection, we assign each device $n$ a unique signature sequence set of size $Q \triangleq 2^J$:
\begin{equation}
	\mathcal{S}_n \triangleq \left\{ \bm{s}_{n,1}, \bm{s}_{n,2}, \ldots, \bm{s}_{n,Q} \right\},
\end{equation}
where $\bm{s}_{n,q} \in \mathbb{C}^L,$ $1 \le q \le Q.$
When device $n$ is active and needs to send $J$ bits of data,
it selects a single sequence from its assigned set $\mathcal{S}_n$ and sends this sequence to the BS.
The BS then detects which sequences have been transmitted to perform the device activity detection and data decoding.

The process of performing joint activity and data detection using the covariance-based approach is as follows.
Let $a_{n,q} \in \{0,1\}$ indicate whether sequence $q$ of device $n$ (i.e., $\bm{s}_{n,q}$) is transmitted.
Since at most one sequence is selected by each device, $a_{n,q}$ satisfies $\sum_{q=1}^{Q} a_{n,q} \in \{0,1\}$ where $\sum_{q=1}^{Q} a_{n,q} = 1$
indicates that device $n$ is active, and $\sum_{q=1}^{Q} a_{n,q} = 0$ indicates
that device $n$ is inactive.
Then, the received signal at the BS, as well as its vectorization, can be expressed as follows:
\begin{align}
	\bm{Y} & = \sum_{n=1}^{N} \sum_{q=1}^{Q} a_{n,q}\bm{s}_{n,q}\bm{h}_{n}^T  + \bm{W}, \\
	\bm{y} & = \operatorname{vec}(\bm{Y}) = \sum_{n=1}^{N} \sum_{q=1}^{Q} a_{n,q} \bm{h}_{n} \otimes \bm{s}_{n,q} + \bm{w}.
\end{align}
Therefore, we have $\bm{y}\sim\mathcal{CN}(\overline{\bm{y}}_{\bm{a}},\bm{\Sigma}_{\bm{a}}),$ where the mean is
\begin{equation}
	\overline{\bm{y}}_{\bm{a}} = \mathbb{E} [ \bm{y} ]  = \sum_{n=1}^{N} \sum_{q=1}^{Q} a_{n,q} \overline{\bm{h}}_{n} \otimes \bm{s}_{n,q}, 
\end{equation}
and the covariance matrix is given by
\begin{align*}
	\bm{\Sigma}_{\bm{a}} & = \mathbb{E}[ (\bm{y}-\overline{\bm{y}}_{\bm{a}}) (\bm{y}-\overline{\bm{y}}_{\bm{a}})^H ] \\
	& = \sum_{n=1}^{N} \sum_{q=1}^{Q} a_{n,q} \bm{R}_n \otimes \left(\bm{s}_{n,q} \bm{s}_{n,q}^H\right) + \sigma_w^2 \bm{I}_{LM}.
\end{align*}
Using the same techniques presented in Section~\ref{subsec:prob-form}, the MLE problem can be expressed as
\begin{subequations}
	\begin{alignat}{2}
		&\underset{\bm{a}}{\operatorname{minimize}}    &\quad& \log|\bm{\Sigma}_{\bm{a}}| +   \left(\bm{y}-\overline{\bm{y}}_{\bm{a}}\right)^H \bm{\Sigma}_{\bm{a}}^{-1} \left(\bm{y}-\overline{\bm{y}}_{\bm{a}}\right)\\
		&\operatorname{subject\,to} &      &a_{n,q} \in \{0,1\}, \, 1\le n \le N, \, 1 \le q \le  Q, \\
		& & & \textstyle\sum_{q=1}^{Q} a_{n,q} \in \{0,1\}, \, 1\le n \le N.
	\end{alignat}
\end{subequations}
To make the problem more computationally tractable, we relax the binary constraint $a_{n} \in \{0,1\}$ to $a_{n} \in [0,1]$ as in~\cite{fengler2021non,chen2022phase,wang2023covariance}, and we relax the combination constraint $\sum_{q=1}^{Q} a_{n,q} \in \{0,1\}$ as in~\cite{chen2019covariance}.
Then we obtain the following relaxed problem:
\begin{subequations}\label{eq:mle-data}
	\begin{alignat}{2}
		\quad &\underset{\bm{a}}{\operatorname{minimize}}    &\quad& \log|\bm{\Sigma}_{\bm{a}}| +   \left(\bm{y}-\overline{\bm{y}}_{\bm{a}}\right)^H \bm{\Sigma}_{\bm{a}}^{-1} \left(\bm{y}-\overline{\bm{y}}_{\bm{a}}\right)\\
		&\operatorname{subject\,to} &      &a_{n,q} \in [0,1], \, 1\le n \le N, \, 1 \le q \le  Q.
	\end{alignat}
\end{subequations}
Therefore, by solving problem~\eqref{eq:mle-data} using the exact CD and inexact CD algorithms presented in this paper, we can effectively perform joint activity and data detection.

\section{Simulation Results}
\label{sec:simulation}

We consider an mMTC system with a single cell of radius $500$\,m, where all potential devices are uniformly distributed.
In the simulation, the carrier frequency is set as $3$\,GHz.
The scatterers are uniformly distributed within $200$\,m from the BS, and the intensity attenuation $\sigma_{\ell}^2$ is $1$ for all $\ell.$
The channel gain $\beta_{\ell}$ and $\beta_{n,\ell}$ follow the path-loss model, which is given by $128.1 + 37.6 \log_{10}(d),$ where $d$ is the corresponding distance in km.
We adopt a power control scheme such that $\mathbb{E}[\|\bm{h}_n\|_2^2] = \rho M,$ where the power $\rho$ is set as $-105.1$\,dBm.
The background noise power is set as $-169$\,dBm/Hz over $10$\,MHz.
The signature sequences are generated randomly and uniformly from the discrete set $\qam^L,$ as in~\cite{wang2023covariance}.
Notice that there are enough devices in the near-field region because, as mentioned in Section~\ref{subsec:system}, the Rayleigh distance can be sufficiently large, especially when the number of antennas $M$ is large.

All experiments are implemented in Matlab 2021b on a computer equipped with an Intel Core i7-11700 CPU and 32\,GB of memory.
This is for a fair comparison in computational complexity with existing algorithms.
In practical implementations, more efficient programming languages such as C/C++ would be adopted, and the code would also be optimized for even more efficient execution.

\subsection{Computational Performance of Presented Algorithms}
\label{sec:compare-alg}

In this subsection, we demonstrate the computational performance of the exact and inexact CD algorithms for solving the device activity detection problem~\eqref{eq:mle}.
In the inexact CD algorithm, we set the parameter $\mu$ to $10.$

\begin{table}[t]
	\centering
	\caption{The Proportion of Experiments in Which the Exact CD Algorithm Converges out of $100$ Experiments.}
	\label{table:rn}
	\resizebox{\linewidth}{!}{
		\begin{tabular}{|l|c|c|c|c|c|} 
			\hline
			\makecell{{}\\{}}& $1 \le \bar{\ell}_n \le 8$ & $\bar{\ell}_n = 9$ & $\bar{\ell}_n = 10$ & $\bar{\ell}_n = 11$ & $12 \le \bar{\ell}_n \le 20$  \\ \hline
			\makecell{Setting 1: $M = 128,$ $L = 20,$\\$N = 200,$ $K = 20$}& $100\%$ & $0\%$ & $0\%$ & $0\%$ & $0\%$  \\ \hline
			\makecell{Setting 2: $M = 64,$ $L = 20,$\\$N = 200,$ $K = 20$}& $100\%$ & $100\%$ & $99\%$ & $15\%$ & $0\%$  \\ \hline
			\makecell{Setting 3: $M = 128,$ $L = 10,$\\$N = 200,$ $K = 20$}& $100\%$ & $100\%$ & $97\%$ & $0\%$ & $0\%$  \\ \hline
			\makecell{Setting 4: $M = 128,$ $L = 20,$\\$N = 100,$ $K = 20$}& $100\%$ & $0\%$ & $0\%$ & $0\%$ & $0\%$  \\ \hline
			\makecell{Setting 5: $M = 128,$ $L = 20,$\\$N = 200,$ $K = 10$}& $100\%$ & $0\%$ & $0\%$ & $0\%$ & $0\%$  \\ \hline
	\end{tabular}}
\end{table}

Table~\ref{table:rn} shows the convergence behavior of the exact CD algorithm for different $\bar{\ell}_n$'s. For each $\bar{\ell}_n,$ we record the proportion of experiments in which the exact CD algorithm converges out of $100$ experiments.
We observe that for the system parameter settings listed in Table~\ref{table:rn}, the exact CD
algorithm always converges when $\bar{\ell}_n \le 8$ (corresponding to $r_n \le 8$).
Conversely, the algorithm always diverges when $\bar{\ell}_n \ge 12$ (for most cases $r_n \ge 12$).
Comparing Setting~1 with Setting~2, we observe that as the number of antennas $M$ decreases from $128$ to $64,$ the convergence threshold for $\bar{\ell}_n$ increases from $8$ to $10.$
Similarly, comparing Setting~1 and Setting~3, we find that reducing the signature sequence length from $20$ to $10$ also increases the convergence threshold for $\bar{\ell}_n$ from $8$ to $10.$
Furthermore, by comparing Setting~1 with Settings~4 and~5, we observe that parameters $N$ and $K$ do not influence the convergence threshold of $\bar{\ell}_n.$

\begin{figure}[t]
	\centering
	\begin{subfigure}[t]{0.49\columnwidth}
		\includegraphics[width=\textwidth]{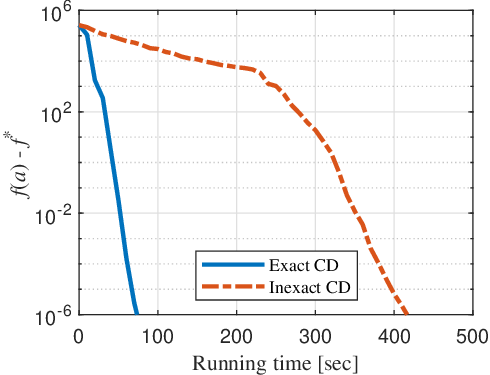}
		\caption{$\bar{\ell}_n = 2$ for all $n.$}
		\label{fig:alg-2}
	\end{subfigure}
	\hfill
	\begin{subfigure}[t]{0.49\columnwidth}
		\includegraphics[width=\textwidth]{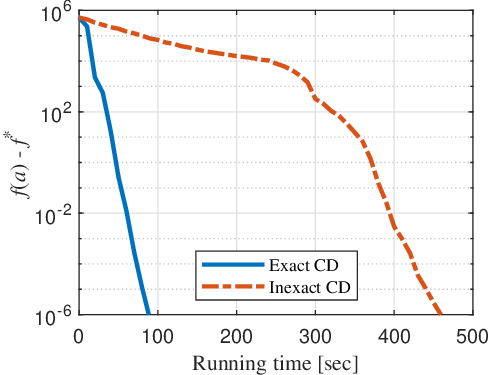}
		\caption{$\bar{\ell}_n = 4$ for all $n.$}
		\label{fig:alg-4}
	\end{subfigure}
	\hfill
	\begin{subfigure}[t]{0.49\columnwidth}
		\includegraphics[width=\textwidth]{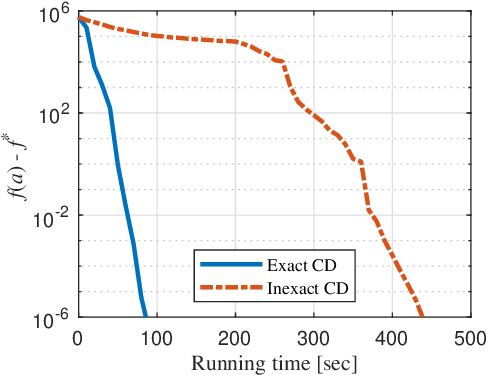}
		\caption{$\bar{\ell}_n = 8$ for all $n.$}
		\label{fig:alg-8}
	\end{subfigure}
	\hfill
	\begin{subfigure}[t]{0.49\columnwidth}
		\includegraphics[width=\textwidth]{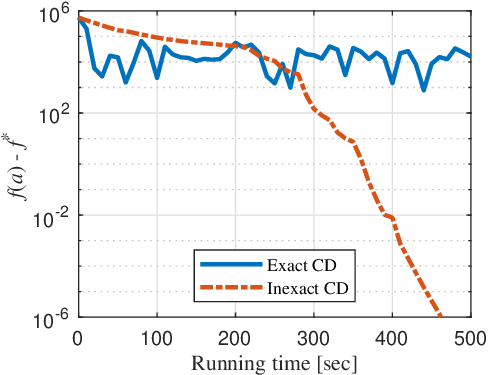}
		\caption{$\bar{\ell}_n = 12$ for all $n.$}
		\label{fig:alg-12}
	\end{subfigure}
	\caption{Comparison of the objective function value of the presented exact and inexact CD algorithms versus the running time.}
	\label{fig:alg}
\end{figure}

In Fig.~\ref{fig:alg}, we show the decrease of the objective function values of the exact and inexact CD algorithms with respect to the running time with $M = 128,$ $L = 20,$ $N = 200,$ and $K = 20.$
The outputs of the exact and inexact CD algorithms are denoted as $\bm{a}^{(e)}$ and $\bm{a}^{(i)},$ respectively.
We then define $f^*$ as the minimum of $f(\bm{a}^{(e)})$ and $f(\bm{a}^{(i)}).$
We observe from Fig.~\ref{fig:alg} that when $\bar{\ell}_n = 2,4,$ and $8,$ the objective function values of exact and inexact CD can gradually decrease and converge to the same value.
Moreover, exact CD is significantly more efficient than inexact CD.
This is because, when $\bar{\ell}_n$ is small (implying $r_n$ is also small), the exact CD algorithm can find the exact solution to the one-dimensional optimization subproblem~\eqref{eq:one-dim-r} in each coordinate update, whereas the inexact CD algorithm can only find an approximate solution.
In Fig.~\ref{fig:alg-12}, we observe that when $\bar{\ell}_n = 12,$ the objective function value of the inexact CD algorithm still converges, but that of the exact CD algorithm cannot decrease and diverges.
This observation shows that exact CD is more efficient when $\bar{\ell}_n$ is small, while inexact CD is more robust when $\bar{\ell}_n$ is large.

\begin{figure}[t]
	\centering
	\begin{subfigure}[t]{0.49\columnwidth}
		\includegraphics[width=\textwidth]{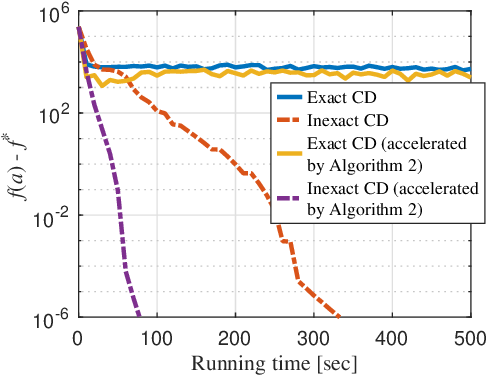}
		\caption{$N_{\mathrm{corr}} = 20,$ $N = 200,$ and $\bar{\ell}_n = 4$ for $1 \le n \le N_{\mathrm{corr}}.$}
		\label{fig:alg-acc-a}
	\end{subfigure}
	\hfill
	\begin{subfigure}[t]{0.49\columnwidth}
		\includegraphics[width=\textwidth]{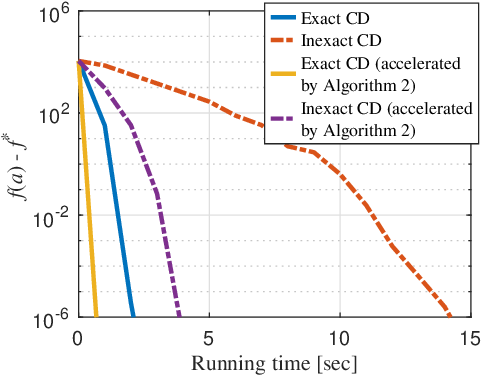}
		\caption{$N_{\mathrm{corr}} = 20,$ $N = 20,$ and $\bar{\ell}_n = 4$ for all $n.$}
		\label{fig:alg-acc-b}
	\end{subfigure}
	\caption{Comparison of the objective function value of the presented exact and inexact CD algorithms as well as their accelerated versions by Algorithm~\ref{alg:update-sigma} versus the running time.}
	\label{fig:alg-acc}
\end{figure}

In Fig.~\ref{fig:alg-acc}, we assume that Assumption~\ref{assu:rank-deficient} holds true and plot the decrease of the objective function values of the exact and inexact CD algorithms as well as their accelerated versions by Algorithm~\ref{alg:update-sigma} with respect to the running time with $M = 128$ and $L = 20.$
In Fig.~\ref{fig:alg-acc-a}, we set $N = 200,$ $N_{\mathrm{corr}} = 20$ in Assumption~\ref{assu:rank-deficient}, and $\bar{\ell}_n = 4$ for $1 \le n \le N_{\mathrm{corr}}$ (implying that $r_n \le 4$).
Consequently, we have $r^{\prime} \le 80 < M$ in \eqref{eq:sum-rank-deficient}. 
Notice that in this case, there are $180$ devices with channel correlation matrix $\bm{R}_n = g_n \bm{I}_M.$
Fig.~\ref{fig:alg-acc-a} shows that, in this scenario, 
the objective function value of the accelerated inexact CD converges to $f^*,$ and it is more computationally efficient than the inexact CD algorithm.
However, neither exact CD nor its accelerated version converges.
This is because neither exact CD nor its accelerated version can solve the one-dimensional optimization subproblem~\eqref{eq:one-dim-r} when the channel correlation matrix of the device $\bm{R}_n = g_n \bm{I}_M.$
To demonstrate that the exact CD algorithm can also be accelerated when it converges,
we remove devices with channel correlation matrix $\bm{R}_n = g_n \bm{I}_M.$
Therefore, in Fig.~\ref{fig:alg-acc-b}, we set $N = 20,$ $N_{\mathrm{corr}} = 20$ in Assumption~\ref{assu:rank-deficient}, and $\bar{\ell}_n = 4$ for all $n.$
We observe from Fig.~\ref{fig:alg-acc-b} that the accelerated version of exact CD converges, and it is more efficient than exact CD.
The same observation applies to the inexact CD algorithm as well.

A summary of the presented CD algorithms is as follows.
First, the exact CD algorithm is more computationally efficient than the inexact CD algorithm when the rank of the channel correlation matrix is small.
Second, the inexact CD algorithm is more favorable in terms of robustness in the case where the rank of the channel correlation matrix is large.
Finally, both of the exact and inexact CD algorithms can be accelerated when Assumption~\ref{assu:rank-deficient} holds. Due to its robustness in solving the subproblem, we adopt the inexact CD algorithm to solve problem~\eqref{eq:mle} in different scenarios in the following two subsections.

\subsection{Numerical Validation of Detection Performance Analysis}

In this subsection, we verify the detection performance analysis in Section~\ref{sec:analysis}.
We consider the device activity detection problem under Case~\ref{item:corr} and Case~\ref{item:uncorr} and perform MLE model~\eqref{eq:mle} in both cases.
Notice that the key difference between Case~\ref{item:corr} and Case~\ref{item:uncorr} lies in the channel correlation matrices.
In order to highlight the impact of the channel correlation matrices on detection performance while eliminating the effect of channel means, we set the channel means to zero in this subsection.
The termination condition of the inexact CD algorithm is set as $\|\V(\bm{a})\|_2 \le 10^{-3}.$

\begin{figure}[t]
	\centering
	\includegraphics[width=0.76\columnwidth]{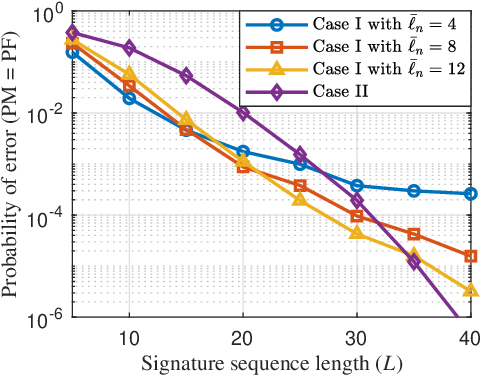}
	\caption{Comparison of detection performance between Case~\ref{item:corr} and Case~\ref{item:uncorr} for different $L$'s.}
	\label{fig:compare-cases-L}
\end{figure}

In Fig.~\ref{fig:compare-cases-L}, we show the detection performance for different values of $L$ with $M = 32,$ $N = 200,$ and $K = 20.$
The ``probability of error'' in the figure is defined as the point where the probability of missed detection (PM) and probability of false alarm (PF) are equal.
We observe that increasing $L$ can improve the detection performance in both cases.
When $L$ is small, the detection performance in Case~\ref{item:corr} is better, and when $L$ is large, the detection performance in Case~\ref{item:uncorr} is better.
This is because $\bm{R}_n$ is rank-deficient in Case~\ref{item:corr} and full-rank in Case~\ref{item:uncorr}, which makes the signal statistical dimension (defined in \eqref{eq:def-D-I} and \eqref{eq:def-D-II}) in Case~\ref{item:uncorr} larger than that in Case~\ref{item:corr} when $L$ is large.
In other words, in Case~\ref{item:corr}, some dimensions of the received signal are actually useless.
We can also observe from Fig.~\ref{fig:compare-cases-L} that in Case~\ref{item:corr}, when $L \ge 35,$ the larger $\bar{\ell}_n$ is, the better the detection performance is.
This also verifies when $\bar{\ell}_n$ is larger, the rank of $\bm{R}_n$ is larger, and thus the signal statistical dimension is larger, and the detection performance is better.
The crossover for $\bar{\ell}_n=8$ and $\bar{\ell}_n=12$ is because when $L$ is small, the signal statistical dimension is the same for both cases, as it cannot exceed $LM.$
When $L$ is large, the larger the $\bar{\ell}_n$ is, the larger the rank of $\bm{R}_n$ will be, and consequently, the larger the signal statistical dimension.
Therefore, the detection performance for $\bar{\ell}_n=12$ is better than that for $\bar{\ell}_n=8.$

\begin{figure}[t]
	\centering
	\includegraphics[width=0.76\columnwidth]{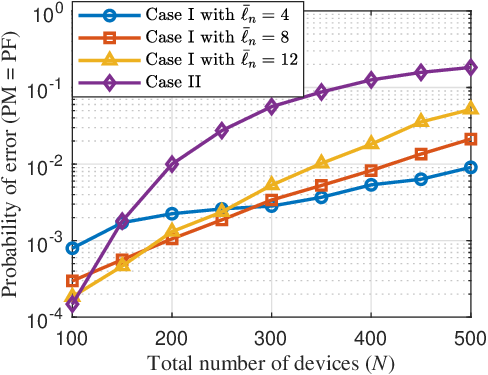}
	\caption{Comparison of detection performance between Case~\ref{item:corr} and Case~\ref{item:uncorr} for different $N$'s.}
	\label{fig:compare-cases-N}
\end{figure}

Fig.~\ref{fig:compare-cases-N} plots the detection performance for various values of $N,$ with parameters $M = 32,$ $L = 20,$ and $K = N/10.$
It is observed that the detection performance of both cases deteriorates as $N$ increases.
Furthermore, when $N$ is large, for instance, $N \ge 150,$ the detection performance in Case~\ref{item:corr} is superior to that of Case~\ref{item:uncorr}.
This is due to when $N$ is large, the signal statistical dimensions are equal in both cases; however, in Case~\ref{item:corr}, the true activity vector is more easily identifiable compared to Case~\ref{item:uncorr}, as discussed in Section~\ref{subsec:id}.
Additionally, we also observe that in Case~\ref{item:corr}, a smaller $\bar{\ell}_n$ tends to result in better detection performance for larger $N.$
This can be explained by the fact that when $\bar{\ell}_n$ is smaller, the cosine similarity in \eqref{eq:cosine} is likely to be lower, indicating a weaker interference between different devices, which in turn leads to improved detection performance.

\begin{figure}[t]
	\centering
	\includegraphics[width=0.76\columnwidth]{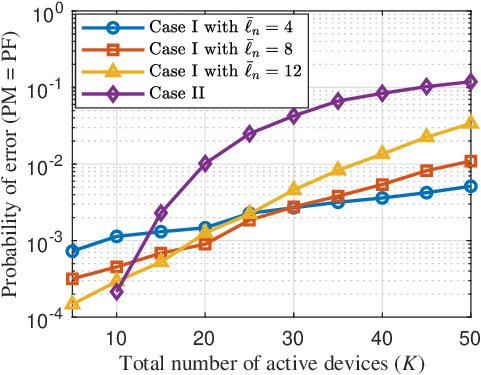}
	\caption{Comparison of detection performance between Case~\ref{item:corr} and Case~\ref{item:uncorr} for different $K$'s.}
	\label{fig:compare-cases-K}
\end{figure}

Fig.~\ref{fig:compare-cases-K} plots the detection performance for various values of $K,$ with parameters $M = 32,$ $L = 20,$ and $N = 200.$
Notice that the sparsity level of active devices ranges from $0.025$ to $0.25.$
We observe that the detection performance in both cases deteriorates as $K$ increases, exhibiting a similar trend to that as in Fig.~\ref{fig:compare-cases-N} when $N$ increases.
When $K$ is large, the detection performance in Case~\ref{item:corr} is superior to that in Case~\ref{item:uncorr}.
Additionally, within Case~\ref{item:corr}, a smaller $\bar{\ell}_n$ value corresponds to better detection performance for large $K.$
This is because, as discussed in Section~\ref{subsec:id}, the interference between devices in Case~\ref{item:corr} is less than that in Case~\ref{item:uncorr}.
Moreover, in Case~\ref{item:corr}, a smaller $\bar{\ell}_n$ value corresponds to a smaller cosine similarity in \eqref{eq:cosine}, resulting in less interference between devices.

\begin{figure}[t]
	\centering
	\includegraphics[width=0.76\columnwidth]{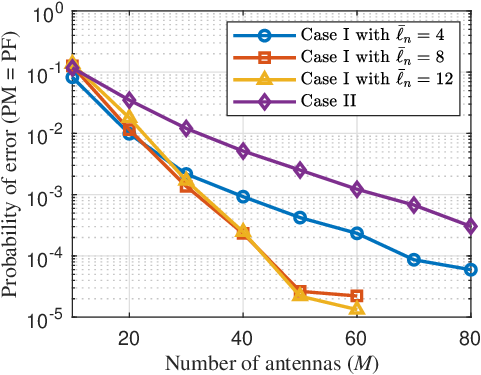}
	\caption{Comparison of detection performance between Case~\ref{item:corr} and Case~\ref{item:uncorr} for different $M$'s.}
	\label{fig:compare-cases-M}
\end{figure}

Fig.~\ref{fig:compare-cases-M} shows the detection performance for various values of $M,$ with $L = 20,$ $N = 200,$ and $K = 20.$
We observe that as $M$ increases, the detection performance in Case~\ref{item:corr} with $\bar{\ell}_n$ values of $8$ and $12$ significantly outperforms that of Case~\ref{item:corr} with $\bar{\ell}_n = 4$ and Case~\ref{item:uncorr}.
The detection performance of Case~\ref{item:corr} with $\bar{\ell}_n=4$ is worse than that of Case~\ref{item:corr} with $\bar{\ell}_n=8$ and $\bar{\ell}_n=12$ when $M$ is large, because the signal statistical dimension in Case~\ref{item:corr} with $\bar{\ell}_n=4$ is lower than that in Case~\ref{item:corr} with $\bar{\ell}_n=8$ and $\bar{\ell}_n=12.$
The detection performance of Case~\ref{item:uncorr} is worse than that of Case~\ref{item:corr} with $\bar{\ell}_n=8$ and $\bar{\ell}_n=12$ because, in Case~\ref{item:corr}, the identifiability of the true activity vector $\bm{a}^{\circ}$ is easier to hold, and there is less mutual interference between different devices.

\subsection{Detection Performance Comparison with Benchmarks}

In this subsection, we compare the detection performance of the proposed MLE-based approach with the following benchmarks:
\begin{itemize}
	\item AMP~\cite{djelouat2021user} and OAMP~\cite{cheng2021orthogonal}:	
	Both algorithms are designed for joint activity detection and channel estimation under correlated Rayleigh fading channels.
	OAMP is an improved version of AMP, which is applicable to more types of signature sequences.
	Notice that while AMP and OAMP can be applied under correlated Rayleigh fading channels, they cannot be directly applied to correlated Rician fading.
	The reason is that the denoiser and state evolution in these algorithms assume zero channel means, which cannot be directly extended to scenarios with nonzero channel means.
	In the scenario of Rician fading channels, i.e., $\bm{h}_n \sim \mathcal{CN}\big( \overline{\bm{h}}_n, \bm{R}_n \big),$ we ignore the mean $\overline{\bm{h}}_n$ and assume the prior distribution to be $\bm{h}_n \sim \mathcal{CN}\big( \bm{0}, \bm{R}_n \big)$ in both methods.
	
	Notice that both methods require the knowledge of the probability that a device is active, i.e., $K/N,$ but the MLE does not require this prior information.
	\item Mismatched MLE:
	We apply the vanilla MLE-based approach in \cite{liu2023mle}.
	Specifically, we use $\bm{R}_n^{\prime} = \frac{\mathrm{tr}(\bm{R}_n)}{M} \bm{I}_M$ to replace the channel correlation matrix $\bm{R}_n$ in the MLE formulation.
\end{itemize}
The termination condition of the inexact CD algorithm is set as $\|\V(\bm{a})\|_2 \le 10^{-3},$ and we allow other benchmarks to run sufficiently long until they converge.

Notice that AMP~\cite{djelouat2021user} and OAMP~\cite{cheng2021orthogonal} are CS-based methods, while the MLE proposed in this paper and mismatched MLE are covariance-based methods.
The key difference between the CS-based method and the covariance-based method is that the former jointly estimates the channels and device activities, whereas the latter estimates only the device activities.
Other covariance-based methods include SPICE~\cite{stoica2011spice}, which can be applied to the Rayleigh fading scenario but encounters technical difficulties in the Rician fading scenario.

\begin{figure}[t]
	\centering
	\begin{subfigure}[t]{0.49\columnwidth}
		\includegraphics[width=\textwidth]{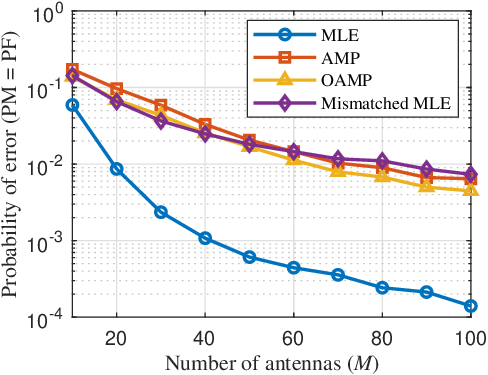}
		\caption{Detection under correlated Rician fading channels.}
		\label{fig:compare-M-a}
	\end{subfigure}
	\hfill
	\begin{subfigure}[t]{0.49\columnwidth}
		\includegraphics[width=\textwidth]{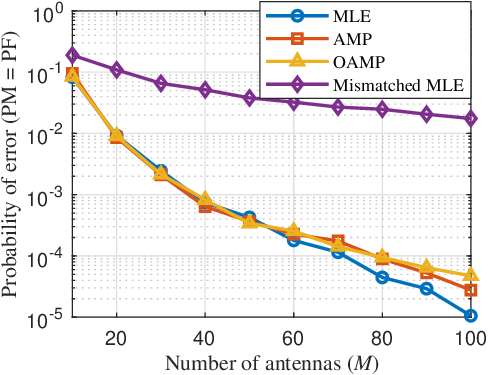}
		\caption{Detection under correlated Rayleigh fading channels.}
		\label{fig:compare-M-b}
	\end{subfigure}
	\caption{Detection performance comparison of the MLE approach and the benchmarks for different $M$'s.}
	\label{fig:compare-M}
\end{figure}

In Fig.~\ref{fig:compare-M}, we show the detection performance under correlated Rician/Rayleigh fading channels for various values of $M,$ with $L = 20,$ $N = 200,$ $K = 20,$ and $\bar{\ell}_n = 4.$
It can be observed from Fig.~\ref{fig:compare-M} that increasing $M$ can improve the detection performance of all compared methods.
In Fig.~\ref{fig:compare-M-a}, we observe that the detection performance of MLE is significantly better than that of all the other benchmarks.
The main reason for this is that these benchmarks use inaccurate channel distributions: both AMP and OAMP do not utilize the LoS channel, and the mismatched MLE roughly approximates $\bm{R}_n.$
In Fig.~\ref{fig:compare-M-b}, we observe that the detection performance of MLE is slightly better than that of AMP and OAMP when $M$ is large and significantly better than that of mismatched MLE.
In this case, all methods except mismatched MLE use accurate channel distributions. Although the detection performance of MLE, AMP, and OAMP is relatively close, a significant advantage of MLE is that it does not require the prior knowledge of $K.$

\begin{figure}[t]
	\centering
	\begin{subfigure}[t]{0.49\columnwidth}
		\includegraphics[width=\textwidth]{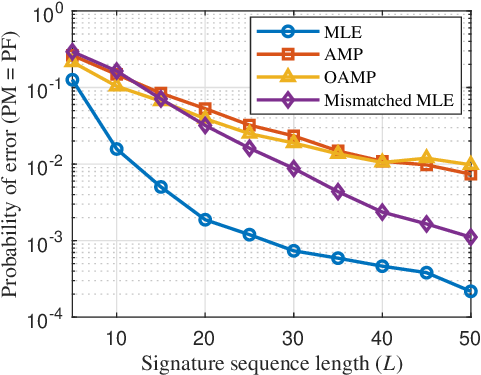}
		\caption{Detection under correlated Rician fading channels.}
		\label{fig:compare-L-a}
	\end{subfigure}
	\hfill
	\begin{subfigure}[t]{0.49\columnwidth}
		\includegraphics[width=\textwidth]{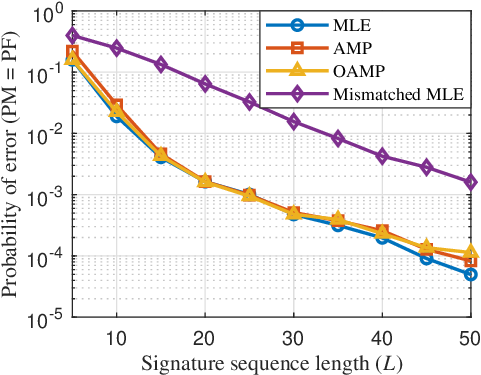}
		\caption{Detection under correlated Rayleigh fading channels.}
		\label{fig:compare-L-b}
	\end{subfigure}
	\caption{Detection performance comparison of the MLE approach and the benchmarks for different $L$'s.}
	\label{fig:compare-L}
\end{figure}

Fig.~\ref{fig:compare-L} plots the probability of error for different values of $L,$ with $M = 32,$ $N = 200,$ $K = 20,$ and $\bar{\ell}_n = 4.$
We observe that increasing $L$ can improve the detection performance of all compared methods.
In Fig.~\ref{fig:compare-L-a}, due to the use of inaccurate channel distributions, the detection performance of AMP, OAMP, and mismatched MLE is significantly worse than that of MLE.
Moreover, when $L$ is large, the detection performance of mismatched MLE becomes better than that of AMP and OAMP.
In Fig.~\ref{fig:compare-L-b}, we observe that the detection performances of MLE, AMP, and OAMP are very close, and the detection performance of mismatched MLE is significantly worse than theirs.

\begin{figure}[t]
	\centering
	\begin{subfigure}[t]{0.49\columnwidth}
		\includegraphics[width=\textwidth]{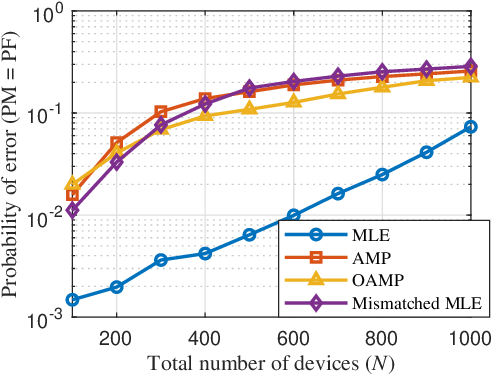}
		\caption{Detection under correlated Rician fading channels.}
		\label{fig:compare-N-a}
	\end{subfigure}
	\hfill
	\begin{subfigure}[t]{0.49\columnwidth}
		\includegraphics[width=\textwidth]{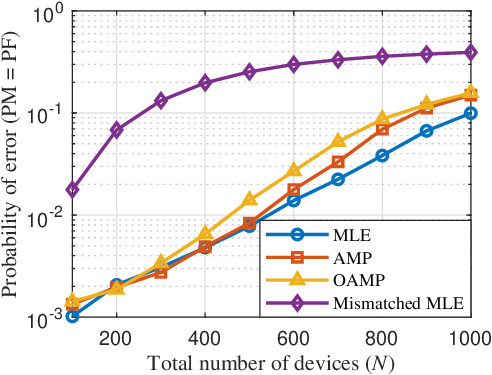}
		\caption{Detection under correlated Rayleigh fading channels.}
		\label{fig:compare-N-b}
	\end{subfigure}
	\caption{Detection performance comparison of the MLE approach and the benchmarks for different $N$'s.}
	\label{fig:compare-N}
\end{figure}

Fig.~\ref{fig:compare-N} plots the probability of error for various values of $N,$ with $M = 32,$ $L = 20,$ $K = N/10,$ and $\bar{\ell}_n = 4.$
We observe that as $N$ increases, the detection performance of the compared methods deteriorates.
Fig.~\ref{fig:compare-N-a} illustrates that in the Rician fading scenario, the detection performance of MLE is significantly better than that of the other methods.
Fig.~\ref{fig:compare-N-b} shows that in the Rayleigh fading scenario, the detection performance of MLE, AMP, and OAMP is very close, with the mismatched MLE performing worse than these methods.
Moreover, MLE outperforms AMP and OAMP when $N$ is large.

\subsection{Joint Device Activity and Data Detection}

In this subsection, we study the joint device activity and data detection problem and demonstrate the detection performance of the MLE model~\eqref{eq:mle-data}.
\begin{figure}[t]
	\centering
	\includegraphics[width=0.76\columnwidth]{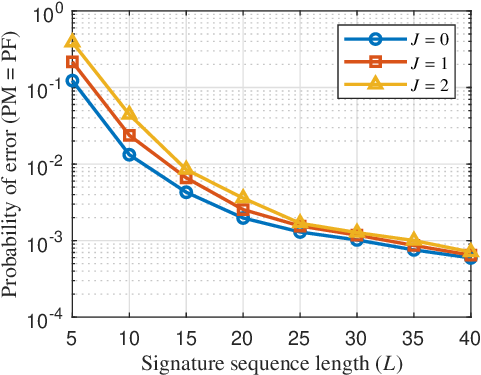}
	\caption{Detection performance comparison of the MLE model in joint device activity and data detection context for different $L$'s.}
	\label{fig:data-detection}
\end{figure}

Fig.~\ref{fig:data-detection} plots the detection performance for various values of $L,$ with $M = 32,$ $N = 200,$ $K = 20,$ and $\bar{\ell}_n = 4.$
We observe from Fig.~\ref{fig:data-detection} that when $L$ is small, the smaller the number of information bits $J,$ the better the detection performance.
In addition, as $L$ increases, for bit $J = 0, 1,$ and $2,$ the detection performance of the MLE model improves and gradually becomes equal.
Note that when $J=0,$ there is no data transmission, and only activity detection is performed.
This indicates that in the joint device activity and data detection problem, transmitting $J$ bits of information (when $J$ is small) does not significantly deteriorate the detection performance of the MLE model, especially when $L$ is large.

\section{Conclusion}
\label{sec:conclusion}

This paper investigates the device activity detection problem for NFC, where the near-field channel can be modeled as correlated Rician fading. The corresponding device activity detection problem is formulated as an MLE problem.
We present the exact CD algorithm and propose the inexact CD algorithm with improved numerical stability for solving the MLE problem.
In addition, an analysis of the detection performance of MLE is provided. Simulation results demonstrate the efficiency of the presented algorithms and verify the correctness of the detection performance analysis.

\appendices

\section{Proof of Theorem~\ref{theorem:objective-decrease}}
\label{sec:proof-objective-decrease}

The outline of the proof is as follows. 
First, we prove that for a sufficiently large $\mu,$ $p_{\mathrm{approx}}(d) + \frac{\mu}{2} d^2$ is an upper bound of $f(\bm{a} + d\,\bm{e}_n).$
We then show that inequality \eqref{eq:objective-decrease} holds by studying the upper bound $p_{\mathrm{approx}}(d) + \frac{\mu}{2} d^2$ in more detail.

According to~\eqref{eq:approx-det} and~\eqref{eq:approx-inv}, we have
\begin{equation}
	f(\bm{a} + d\,\bm{e}_n) = p_{\mathrm{approx}}(d) + o(d).
\end{equation}
The gradient in \eqref{eq:gradient} is derived from~\eqref{eq:p-approx} and
\begin{equation}\label{eq:derivation-gradient}
	\left. [\nabla f(\bm{a})]_n = \frac{\partial f(\bm{a} + d\,\bm{e}_n)}{\partial \, d} \right|_{d = 0} =  p_{\mathrm{approx}}^{\prime}(0).
\end{equation}
Since $\nabla f(\bm{a})$ is continuous and $\bm{a}$ is in the bounded region $[0,1]^N,$ $\nabla f(\bm{a})$ is Lipschitz continuous. That is, there exists a constant $\lip>0$ such that for any $\bm{a}_1, \bm{a}_2 \in [0,1]^N,$ the following inequality holds:
\begin{equation}
	\left\| \nabla f(\bm{a}_1) - \nabla f(\bm{a}_2) \right\|_2 \le \lip \left\| \bm{a}_1 - \bm{a}_2 \right\|_2.
\end{equation}
Furthermore, we have a quadratic upper bound on $f(\bm{a})$ \cite[Proposition~A.24]{bertsekas1999nonlinear}:
\begin{equation}\label{eq:f-upper}
	f(\bm{a}_2) \le f(\bm{a}_1) + \left(\nabla f(\bm{a}_1)\right)^T \left( \bm{a}_2 - \bm{a}_1 \right) + \frac{\lip}{2} \left\| \bm{a}_2 - \bm{a}_1 \right\|^2_2.
\end{equation}
Setting $\bm{a}_1 = \bm{a}$ and $\bm{a}_2 = \bm{a} + d\,\bm{e}_n$ in \eqref{eq:f-upper}, we obtain
\begin{equation}\label{eq:f-upper-d}
	f(\bm{a} + d\,\bm{e}_n) \le f(\bm{a}) + \left[ \nabla f(\bm{a}) \right]_n d + \frac{\lip}{2} d^2.
\end{equation}
Next, we combine all the homogeneous terms in~\eqref{eq:p-approx} and denote $c_{i}(\bm{a})$ as the coefficient of $d^i,$ $i = 2,3,4.$
Specifically,
\begin{multline}
	p_{\mathrm{approx}}(d) = f(\bm{a}) + \\ \left[ \nabla f(\bm{a}) \right]_n d + c_{2}(\bm{a})\, d^2 + c_{3}(\bm{a})\, d^3 + c_{4}(\bm{a})\, d^4.
\end{multline}
Since $c_{2}(\bm{a}),$ $c_{3}(\bm{a}),$ and $c_{4}(\bm{a})$ are continuous and $\bm{a}$ is in the bounded region $[0,1]^N,$ they are bounded.
That is, there exist constants $\bar{c}_{2},$ $\bar{c}_{3},$ and $\bar{c}_{4} > 0$ such that
\begin{equation}
	|c_{i}(\bm{a})| \le \bar{c}_{i}, \quad i = 2,3,4.
\end{equation}
Therefore, setting $c = 2(\bar{c}_{2}+\bar{c}_{3}+\bar{c}_{4}),$ for $\mu \ge \lip + c,$ we have that for any $d \in [-a_{n},\,1-a_{n}],$ the following inequality holds
\begin{equation}\label{eq:approx-upper}
	f(\bm{a}) + \left[ \nabla f(\bm{a}) \right]_n d + \frac{\lip}{2} d^2 \le p_{\mathrm{approx}}(d) + \frac{\mu}{2} d^2.
\end{equation}
Combining~\eqref{eq:f-upper-d} and \eqref{eq:approx-upper}, we obtain
\begin{equation}\label{eq:quad-upper}
	f(\bm{a} + d\,\bm{e}_n) \le p_{\mathrm{approx}}(d) + \frac{\mu}{2} d^2.
\end{equation}

Now we study the optimal value of $p_{\mathrm{approx}}(d) + \frac{\mu}{2} d^2$ to obtain the desired inequality~\eqref{eq:objective-decrease}.
Notice that for any $d \in [-a_{n},\,1-a_{n}],$
\begin{equation}
	p_{\mathrm{approx}}(d) + \frac{\mu}{2} d^2 \le f(\bm{a}) + \left[ \nabla f(\bm{a}) \right]_n d + \frac{\mu + c}{2} d^2.
\end{equation}
Therefore, the optimal value of $p_{\mathrm{approx}}(d) + \frac{\mu}{2} d^2$ satisfies
\begin{align}\label{eq:approx-less}
	& p_{\mathrm{approx}}\big(\bar{d}\big) + \frac{\mu}{2} \big(\bar{d}\big)^2 \nonumber \\
	& \le \underset{d \in [-a_{n},\,1-a_{n}]}{\operatorname{minimize}}~ f(\bm{a}) + \left[ \nabla f(\bm{a}) \right]_n d + \frac{\mu + c}{2} d^2 \nonumber \\
	& \overset{(a)}{\le} f(\bm{a}) 
	- \frac{\left( \operatorname{Proj} \left(a_{n}- \left[\nabla f(\bm{a})\right]_{n}\right) - a_{n} \right)^2}{2(\mu + c)} \nonumber \\
	& = f(\bm{a}) - \frac{[ \V(\bm{a}) ]_n^2}{2(\mu + c)},
\end{align}
where $(a)$ is derived from
\begin{multline}
	\operatorname{Proj} \left(a_{n} - \frac{\left[\nabla f(\bm{a})\right]_{n}}{\mu + c}\right) - a_{n} \\  = \underset{d \in [-a_{n},\,1-a_{n}]}{\arg\min}~ f(\bm{a}) + \left[ \nabla f(\bm{a}) \right]_n d + \frac{\mu + c}{2} d^2
\end{multline}
and the properties of the projection operator $\operatorname{Proj}(\cdot).$
Similar techniques are used in the proof of \cite[Proposition~4]{wang2023covariance}.
Substituting $d = \bar{d}$ into \eqref{eq:quad-upper} and combining it with \eqref{eq:approx-less}, we obtain \eqref{eq:objective-decrease}.
This completes the proof of Theorem~\ref{theorem:objective-decrease}.

\section{Proof of Proposition~\ref{prop:submatrix}}
\label{sec:proof-submatrix}

Let $\bm{u}_m$ denote the $m$-th column vector of $\bm{U}.$ We observe from \eqref{eq:eig-R} that
\begin{equation}\label{eq:sum-u}
	\sum_{n=1}^{N_{\mathrm{corr}}}\bm{u}_m^H \bm{R}_n \bm{u}_m = 0, \quad  r^{\prime} + 1 \le  m \le  M.
\end{equation}
Since $\bm{R}_n$ is positive semidefinite, $\bm{u}_m^H \bm{R}_n \bm{u}_m$ is nonnegative.
Therefore, \eqref{eq:sum-u} implies $\bm{u}_m^H \bm{R}_n \bm{u}_m = 0,$ which is equivalent to $\bm{R}_n \bm{u}_m = \bm{0},$ $r^{\prime} + 1 \le m \le M,$ $1 \le n \le N_{\mathrm{corr}}.$
Now we can verify that only the upper left $r^{\prime} \times r^{\prime}$ submatrix of $\bm{U}^H \bm{R}_n \bm{U}$ is nonzero.
This completes the proof of Proposition~\ref{prop:submatrix}.

\section{Proof of Theorem~\ref{theorem:mle-u}}
\label{sec:proof-mle-u}

According to the property of the Kronecker product, for any matrices $\bm{A}, \bm{B} \in \mathbb{C}^{M \times M}$ and $\bm{C}, \bm{D} \in \mathbb{C}^{L \times L},$ 
\begin{equation}
	\left(\bm{A} \otimes \bm{C}\right) \left(\bm{B} \otimes \bm{D}\right) = \bm{A}\bm{B} \otimes \bm{C}\bm{D}.
\end{equation}
Therefore, we can verify that $(a)$ in \eqref{eq:sigma-u} and $(b)$ in \eqref{eq:y-overline-u} hold.
Furthermore, since $\bm{U}^H \otimes \bm{I}_L \in \mathbb{C}^{LM \times LM}$ is a unitary matrix, i.e., $\big(\bm{U} \otimes \bm{I}_L\big) \big(\bm{U}^H \otimes \bm{I}_L\big) = \bm{I}_{LM},$ we have $\log|\bm{\Sigma}^{\prime}_{\bm{a}}| = \log|\bm{\Sigma}_{\bm{a}}|$ and 
\begin{equation}
	\left(\bm{y}^{\prime}-\overline{\bm{y}}^{\prime}_{\bm{a}}\right)^H \left(\bm{\Sigma}^{\prime}_{\bm{a}}\right)^{-1} \left(\bm{y}^{\prime}-\overline{\bm{y}}^{\prime}_{\bm{a}}\right)
	=
	\left(\bm{y}-\overline{\bm{y}}_{\bm{a}}\right)^H \bm{\Sigma}_{\bm{a}}^{-1} \left(\bm{y}-\overline{\bm{y}}_{\bm{a}}\right).
\end{equation}
This completes the proof of Theorem~\ref{theorem:mle-u}.

\section{Proof of Theorem~\ref{theorem:nc}}
\label{sec:proof-nc}

We prove Theorem~\ref{theorem:nc} by contradiction.
Suppose there exists a nonzero vector $\bm{x} \in \mathcal{N}_{\mathrm{I}} \cap \mathcal{C}.$
Hence we have
\begin{equation}
	\sum_{n=1}^{N} x_n (\bm{R}_n)_{m,m} \bm{s}_{n} \bm{s}_{n}^H = \bm{0}, \quad m = 1,2,\ldots,M.
\end{equation}
Since in Case~\ref{item:uncorr}, $g_n = \frac{1}{M}\mathrm{tr}(\bm{R}_n),$ we have
\begin{equation}
	\sum_{n=1}^{N} x_n g_n \bm{s}_{n} \bm{s}_{n}^H = \frac{1}{M} \sum_{m=1}^{M} \sum_{n=1}^{N} x_n (\bm{R}_n)_{m,m} \bm{s}_{n} \bm{s}_{n}^H = \bm{0},
\end{equation}
which further implies
\begin{equation}
	\sum_{n=1}^{N} x_n g_n\bm{I}_M \otimes \left(\bm{s}_{n} \bm{s}_{n}^H\right) = \bm{I}_M \otimes \left( \sum_{n=1}^{N} x_n g_n \bm{s}_{n} \bm{s}_{n}^H \right) =\bm{0}.
\end{equation}
Therefore, it is easy to verify nonzero $\bm{x} \in \mathcal{N}_{\mathrm{II}} \cap \mathcal{C},$ which contradicts the assumption that $\mathcal{N}_{\mathrm{II}} \cap \mathcal{C} = \{\bm{0}\}.$
This completes the proof of Theorem~\ref{theorem:nc}.

\bibliographystyle{IEEEtran}


\end{document}